\documentclass[12pt,preprint]{aastex}

\newcommand{\invcm}{cm$^{-1}$}
\newcommand{\kms}{km\,s$^{-1}$}
\newcommand{\mic}{\mu \mathrm m}

\slugcomment{In preparation for ApJ, 2005}

\shorttitle{Water in Betelgeuse}
\shortauthors{Ryde et al.}

\begin{document}


\title{Water Vapor on Betelgeuse as Revealed by TEXES High-Resolution $12\,\mic$ Spectra}


\author{N. Ryde\altaffilmark{1}}
\affil{Groupe de Recherche en Astronomie et
Astrophysique du Languedoc (GRAAL), Universit\'e Montpellier II,
Montpellier, France}

\email{ryde@astro.uu.se}

\author{G. M. Harper}
\affil{Center for Astrophysics and Space Astronomy - Astrophysics Research Lab, 593 UCB, University of Colorado, Boulder, CO 80309-0593}
\email{gmh@casa.colorado.edu}

\author{M. J. Richter\altaffilmark{2}}
\affil{Department of Physics, University of California at Davis, CA 95616}
\email{richter@physics.ucdavis.edu}

\author{T. K. Greathouse\altaffilmark{2}}
\affil{Lunar and Planetary Institute,
Houston, TX 77058-1113}
\email{greathouse@lpi.usra.edu}

\author{J. H. Lacy}
\affil{Department of Astronomy,
University of Texas at Austin, TX 78712-1083}
\email{lacy@astro.as.utexas.edu}


\altaffiltext{1}{Chercheur associ\'e financed by Centre National de la
Recherche Scientifique (CNRS). On leave from the Department of Astronomy and Space Physics, Uppsala University, Box 515, SE-751\,20 Uppsala, Sweden.}

\altaffiltext{2}{Visiting Astronomer at the Infrared Telescope Facility,
which is operated by the University of Hawaii under Cooperative Agreement
no. NCC 5-538 with the National Aeronautics and Space Administration, Office
of Space Science, Planetary Astronomy Program.}











\begin{abstract}
The outer atmosphere of the M supergiant Betelgeuse is puzzling.
Published observations of different kinds have shed light on different
aspects of the atmosphere, but no unified picture has emerged.
They have shown, for example, evidence of a water envelope (MOLsphere)
that in some studies is found to be optically thick in the mid-infrared.
In this paper, we present high-resolution, mid-infrared spectra
of Betelgeuse recorded with the TEXES spectrograph.
The spectra clearly show absorption features
of water vapor and OH. We show that a spectrum based on a spherical,
hydrostatic model photosphere with
$T_\mathrm{eff} = 3600$ K, an effective
temperature often assumed for Betelgeuse, fails to model the observed lines.
Furthermore, we show that published MOLspheres scenarios
are unable to explain our data.  However, we are able to model
the observed spectrum reasonably well by adopting
a cooler outer photospheric structure corresponding to
$T_\mathrm{mod}=3250$~K.
The success of this model may indicate the observed mid-infrared lines
are formed in cool photospheric surface regions.
Given the uncertainties of the temperature
structure and the likely  presence of
inhomogeneities, we cannot rule out the possibility that our spectrum
could be mostly photospheric, albeit non-classical.
Our data put new, strong constraints on atmospheric models of Betelgeuse
and we conclude
that continued investigation requires
consideration of non-classical model photospheres as well as possible effects
of a MOLsphere.  We show that the mid-infrared
water-vapor features have great diagnostic value for the environments
of K and M (super-) giant star atmospheres.
\end{abstract}



\keywords{stars: individual ($\alpha$ Ori) --
             stars:  atmospheres--
             Infrared: stars}









\section{INTRODUCTION}

The issue of water vapor in the atmosphere of Betelgeuse
\citep[$\alpha$ Orionis; M1-2 Ia-Iab;][]{perkins} has a 40-year history.
A detection was first proposed in 1964 by
\citet{woolf:64}, but this was overlooked until it was
confirmed by \citet{tsuji_2000} through reanalysis
of the original data.
Betelgeuse has
been used as a water-free reference source in the near-infrared, see for example \citet{hinkle:79}.
\cite{antares} were the first, to our knowledge, to have identified
individual rotational lines of water vapor in $R\approx 10,000$
spectra of Betelgeuse.
They modeled their
observations with a plane-parallel, isothermal
layer close
to the location of the onset of the chromospheric temperature
rise.
\citet{tsuji_2000} interpreted the water-vapor signatures
as arising mostly in an outer region in the atmosphere and not
primarily in the photosphere.
\nocite{verhoelst} Verhoelst et al. (2003; 2005) found a discrepancy between
their $T_\mathrm{eff}=3600$\,K photospheric model and
\emph{Infrared Space Observatory (ISO)} spectra of Betelgeuse which
they also attribute to extra water-vapor absorption in a shell
just above the photosphere.
\citet{weiner:03} identified rotational water lines in Betelgeuse in an effort to
determine spectral regions at $11\,$\micron\ which are free from lines and
suitable for interferometric measurements.

Complicating the picture of the outer atmospheres of
red giants and supergiants is the notion of an
extra atmospheric component called the MOLsphere;
a stationary, warm envelope situated above the photosphere but interior
to the cool, expanding circumstellar shell
\citep[see for example][]{tsuji_1997, tsuji_1998,tsuji_2000,matsuura}.
The MOLsphere has been used in several cases
to explain discrepancies between models and
near-infrared observations of late K and M giants and supergiants
showing water vapor \citep[see for
example][]{tsuji_1997,tsuji_1998,yam_99,tsuji_ny,tsuji_2000,tsuji:iau:01}.
The water vapor, at temperatures of $1000$--$2000\,\mbox{K}$
\citep{tsuji_1997}, results in non-photospheric signatures in IR
spectra of M giants.
Indeed, the observed signatures of water vapor in spectra of Betelgeuse by \citet{tsuji_2000} and \citet{verhoelst}
are assumed to originate
from such a layer.

Furthermore, interpretations of interferometric data from supergiants
\citep[see, e.g., ][]{ohnaka:04a,perrin:04a} and giants
\citep[see, e.g., ][]{mennesson:02,tej:03b,ohnaka:04b,weiner:04,perrin:04b}
can also be related to the MOLsphere.
Dynamic computations
qualitatively predict atmospheric layers far from the
photosphere of Mira stars (red giants)
\cite[see, e.g., ][]{woitke:99,tej:03a}, although
the mechanisms by which layers are formed in supergiants are unclear \citep{ohnaka:04a}.
Thus, a paradigm is emerging where molecular `layers'
exist, a fraction of a stellar radius in extent, above the
photosphere.


In this paper, we present
high-resolution ($R=70,000$),
mid-infrared spectra of the wavelength region of $12.19-12.35\,$\micron\ of the supergiant Betelgeuse
that show clear absorption signatures of water vapor and OH at $12\,\mic$.
These data
put new, strong constraints
on the emerging picture of the MOLsphere, at least for supergiants, and need to be reproduced
in future MOLsphere models. We do not attempt to do so in this
paper. Instead we show the importance of modeling the underlying photosphere appropriately. We argue that, by relaxing
the classical assumptions of the outer regions of the supergiant's photosphere,
it might even be possible to interpret our $12\mic$ spectra without an extra atmospheric component,
except for an expected contribution of an optically-thin continuum formed by dust in the
stellar wind.

Pure rotational lines of water vapor in the $11-12\,\mic$ region,
observed using high spectral resolution, mid-infrared spectrographs, are
interesting diagnostics of the atmospheres of giant stars.
The lines should be ubiquitous in spectra of cool stars at wavelengths longer than approximately $10\,\mic$ \citep{antares}
and have been detected in K and M giants as early as spectral type K1
\citep[Arcturus; ][]{ryde:02,ryde:IAU210:H2O,ryde:03,ryde:livermore}.
The usefulness of the lines is connected with their relatively large transition probabilities and their location in an
uncrowded part of the spectrum.
Rotational lines of water vapor at $12\,\mic$ are analysed for Betelgeuse in this paper.

\section{OBSERVATIONS AND DATA REDUCTION}

We have observed and analysed unique, high-resolution spectra of
Betelgeuse at $12\,\micron$.
We  used the Texas Echelon-Cross-Echelle Spectrograph
\citep[TEXES;][]{texes}, a visitor instrument at the
NASA Infrared Telescope Facility (IRTF).
The TEXES spectra are a significant improvement over those obtained two decades ago.
The spectra presented here
were observed on 4 February 2001 with a
total time spent on source of just over 2 minutes.
We observed the $809.8-820.4\,\mbox{cm$^{-1}$}$ or $12.19-12.35\,\mic$  region, which is
roughly the same as presented by \citet{antares}.

We used the TEXES ``hi-lo'' mode, which features a high-resolution grating
to obtain a spectral resolving power of $R=70,000$ at 12~$\mu$m
cross-dispersed by a first-order grating.
In `hi-lo' mode, a single spectral setting covers over 0.2~$\mu$m
with small gaps between orders at the cost of a short, 2.5\arcsec, slit.
The slit width was 1.5\arcsec.

We calibrated and reduced the data according to standard procedures
described in~\citet{texes}.
We nodded the telescope every 5 seconds to remove
background emission from the sky, telescope, and instrument.
Because of the short slit, we
had to nod the target off the slit, but weather conditions were
good enough that we felt this would not introduce systematic sky
noise in the continuum level.  Later examination of individual nod pairs
show no evidence for systematic variations.
We used spectral features in the telluric atmosphere
to determine the wavelength scale.
The spectra were extracted using the spatial profile along the slit.
A comparison of continuum-normalized spectra with different extraction
parameters shows that the line depths vary by less than 1\%\ of the continuum for any
reasonable range of the extraction parameters. The signal-to-noise ratio is high, of the order of 140:1.

Our flat-fielding procedure leaves broad wiggles in the spectra.  The
standard procedure to remove such wiggles is to divide by the spectrum
of a featureless continuum source.  Unfortunately, no such source
exists with a flux comparable to that of Betelgeuse.  Instead,
to each order we fit a 4$^{th}$ order polynomial which is then
used to normalize the spectra.  Spectral regions with
stellar and telluric features were given low weight in the fit.
The normalization procedure introduces uncertainties: the
equivalent width of a given line depends on the determination of
the local continuum and could be particularly uncertain if the line lies close to,
 or partly in, a gap between the orders. This is not unusual due to the relatively
short wavelength ranges of
the orders.  To investigate these uncertainties, we repeatedly inserted
artificial absorption lines, with equivalent widths and FWHMs similar to
detected lines, and processed the data.
Comparing the recovered equivalent widths with the input values, we can
estimate the uncertainty associated with the normalization.  In general, we find that
the continuum fitting systematically results in smaller equivalent widths
at the 10\%\ level.  For roughly half of our synthetic lines, the equivalent
widths recovered were in error, generally reduced, by 15-20\%.
In approximately one quarter of our tests, the equivalent widths determined were in error
by 50\%\ or more.

The high-resolution spectra of Betelgeuse in the $12\,\micron$ region, shown in
our figures, are in the heliocentric radial velocity frame.
We find that the heliocentric radial velocity for the features in our spectra
is approximately 18 \kms. Looking carefully at our wavelength calibration, we
find the solution to be accurate to $0.3~$\kms\ when compared with known
telluric features in the
observed spectral region.  Comparing OH and H$_2$O line
frequencies identified in the data with an electronic version of the
sunspot atlas \citep{solspectrum}, we found agreement to within $0.6~$\kms.

Betelgeuse's photospheric radial velocity is known to be variable, with both ordered
and random components.
\citet{jones28} found a period of 5.78 years which appeared to be present during
1977-1979 \citep{goldberg79}. The systemic radial velocity of the photosphere
is probably representative
of the radial velocity of Betelgeuse's center-of-mass.
\citet{jones28} and \citet{sanford33} found $+21.05\,$\kms\ and $+20.33\,$\kms\ ,
respectively. The centroid radial
velocity of circumstellar emission lines also provides an independent
estimate of the center-of-mass radial velocity. \citet{huggins87} found
$+19.6\pm 0.4\,$\kms\ for the CO(2-1) 1.3\,mm emission line.

The 5.78 year radial velocity variations have variable, full amplitudes
with $\Delta v \le 6\,$\kms\ which are not always apparent \citep{goldberg84}.
During 1985-1989 \citet{smith:89}
confirmed the $420$\,day period of the optical and
ultraviolet continuum, and the chromospheric \ion{Mg}{2} h \& k flux modulations
found by \citet{dupree87} and
\citet{dupree90} from 1984.0-1990.5. During this time the \citet{sanford33}
radial velocity variations are not apparent.
In summary, the radial velocity of the molecular absorption features in our data
are blue-shifted by approximately $2\,$\kms\ with respect to Betelgeuse's center-of-mass,
but not necessarily with respect to its photosphere, at the time of our observations.

Our mid-infrared lines probe the outer photosphere, and we may be detecting
a slight outflow, similar to that detected in \ion{Fe}{1} UV absorption features
by \cite{carpenter}.
Unfortunately, we do not
know the deep photospheric velocity, which could have been determined by, for example,
optical Fe I and Ti I lines at the time of the observations, and hence do not know
the relative velocity of the mid-infrared lines.


\section{MODEL PHOTOSPHERES AND THE GENERATION OF SYNTHETIC SPECTRA}
\label{models}
For the purpose of analyzing our observations, we have generated
synthetic spectra based on model photospheres calculated with
the {\sc marcs} code \citep{marcs:03}. The version of the {\sc
marcs} code used here is the final major update of the code and its input
data in the suite of {\sc marcs} model-photosphere programs first
developed by \citet{marcs:75} and continually
improved since then \citep[see for example ][]{plez:92,jorg:92,BDP:93}. The new models will be fully
described in a series of forthcoming papers in A\&A (Edvardsson et al., Eriksson et al., Gustafsson et
al., J\o rgensen et al., and Plez et al., all in preparation).

The {\sc marcs} hydrostatic, spherical model photospheres are
computed on the assumptions of Local Thermodynamic Equilibrium
(LTE), chemical equilibrium, homogeneity and the
conservation of the total flux (radiative plus convective; the
convective flux being computed using the mixing length
prescription). The  radiation field used in the model generation is
calculated with absorption from atoms and molecules by opacity
sampling at approximately 95\,000
wavelength points over the wavelength range $1300\,\mbox{\AA} $--$
20\,\mbox{$\mu$m}$. The models are calculated with 56 depth points from
a Rosseland optical depth of $\log \tau_\mathrm{Ross}=2.0$ out to $\log
\tau_\mathrm{Ross}=-5.0$, which in our case corresponds to an
optical depth evaluated at $5000\,\mbox{\AA }$ of $\log
\tau_{5000\AA}=-4.1$. The physical height above the $\log
\tau_\mathrm{Ross}=0$ layer of this outermost point is
$2.8\times10^{12}\,\mbox{cm}$ or $6\%$ of the stellar radius.

Data on absorption by atomic species are collected from the
VALD database \citep{VALD} and Kurucz (1995, private
communication). The opacity of CO, CN, CH, OH, NH, TiO, VO, ZrO,
H$_2$O, FeH, CaH, C$_2$, MgH, SiH, and SiO are included and
up-to-date dissociation energies and partition functions are used.

The star's fundamental parameters are needed
as input for the model photosphere calculation.
As discussed, for example, in \citet{lambert:84},
the accuracy of abundances derived from the synthetic spectra
are determined by \emph{(i)} the basic molecular
data, \emph{(ii)} the quality of the spectra, and \emph{(iii)} the defining parameters and assumptions
made for the model atmosphere. The latter is the most problematic, given the quality of data now achievable in the mid-infrared.

Betelgeuse's fundamental stellar parameters are by no means safely determined.
For example, the effective temperatures to be found in the literature vary widely \citep{harper:01,freytag:02}.
\citet{harper:01} point to several reasons for these uncertainties in their detailed discussion of Betelgeuse's
parameters. The most problematic is the angular diameter, which is needed to determine
the stellar radius, but also provides a way to determine its effective
temperature  \citep[see the discussion in][]{harper:01}.
\cite{bester:96} find an effective temperature
of $T_\mathrm{eff}=3290$~K based on size measurements at $11\,$\micron, but they also
conclude that no single temperature and diameter can fit all their data at once.
They argue that the mid-infrared is
the optimal wavelength region in which to measure sizes of late-type stars.
The lower estimates of $T_\mathrm{eff}$ arise primarily from the adoption of large angular
diameters rather than differences in the bolometric luminosity.

Spectroscopic analyses, on the other hand, indicate higher temperatures. Based on the infrared flux method
\citet{lambert:84} derive an effective temperature of $3800\pm100\,$K but discuss the
possibility of lower temperatures.
\citet{lobel:00} discuss effective-temperature variations and different spectroscopic
temperature determinations. From spectral synthesis calculations, they arrive at a Betelgeuse model with a temperature of
$T_\mathrm{eff}=3500$~K.
Furthermore, they find $\log g=-0.5$, solar metallicity, and
broadening due to macroturbulence and rotation modeled
together with a Gaussian profile with a dispersion of  $v_\mathrm{D}=12\,$\kms.
Effective temperatures of about 3500 K  are commonly adopted \citep{bester:96,antares}.
Levesque et al. (2005; submitted to ApJ) suggest the effective
temperature scale of red supergiants
is significantly \emph{warmer} than previously found, putting the M2 spectral type at $3660\,$K.
In this scale they put Betelgeuse's effective temperature at $3650\,$K. Thus, a large span of effective
temperatures exist in the literature for Betelgeuse.

Three further factors complicate the picture of Betelgeuse's photosphere and its defining parameters.
First, Betelgeuse's visual magnitude is variable, a fact that reflects a change in temperature
rather than size \citep{bester:96,morgan:97}.
Second, the photosphere of
Betelgeuse does not  appear homogeneous.
Homogeneity is normally assumed in photospheric modeling.
Observations show complicated asymmetries  that can be modeled
with temporally varying hot spots, see for example \citet{wilson:97}, \citet{tuthill:97}, \citet{dupree:95},
\citet{dupree:96}, and \citet{young:00}.
These hot-spots may be interpreted
as convective hot-spots \citep{svart,wilson:97,dupree:95} or magnetic activity, pulsations, shock structures, or
changing continuous opacity \citep{dupree:96,gray:00}.
Non-static, hydrodynamic 3D models of red supergiants, which allow for the
formation of spatial inhomogeneities due to convection \citep[c.f ][]{freytag:02,freytag:03,ludwig:03},
are still under development, but have already yielded
important qualitative results,  such as revealing convective hot-spots \cite[see for example][]{freytag:IAU03}. Note,
however, that \citet{gray:01} find no spectroscopic evidence
of giant convective cells on the surface of Betelgeuse. Dynamic models are
crucial for the understanding of these types of stars; such models
are physically much more realistic
and their structures depart markedly from static models. Until realistic models are readily available and
the cause and nature of the inhomogeneities is identified, it may be instructive to think in terms of simple, two-component
models
with  a warm and a cold component with  an arbitrary filling factor  to crudely represent
columns of warm rising and cool sinking gas, cf. \citet{svart} and \citet{tsuji_2000}. Cool areas will easily
produce water-vapor opacity \citep{tsuji_2000}.

Thus, the supposed inhomogeneity of the photosphere of Betelgeuse could make it difficult to
calculate a global spectrum based on a classical model photosphere given by one
effective temperature.
Furthermore, while the continuum intensity in the
Rayleigh-Jeans regime is less sensitive to temperature variations and
so one may expect the effects of effective temperature uncertainties
and surface inhomogeneities on line strengths to be smaller in the
infrared \citep[see, for example,][]{ryde:05_munchen}, formation of
water is very sensitive to the change of temperature through the
photosphere.

The third factor is the uncertainties in the outer
photospheric structure.  Mid-infrared lines are in general
formed further out in the atmosphere than are, for example,
near-infrared lines, as a result of the increased continuous opacity.
An important aspect when discussing modeling of the
atmospheric structure (and especially of the outer regions) of cool stars, is the validity of the assumption of
LTE and molecular equilibrium.
Since the transitions
occur within the electronic ground-state, the assumption of
LTE in  analyzing molecules is
probably valid \citep{hinkle_lambert:75}.
However, the assumption of LTE in the computation of the model structure could
lead to erroneous inferences.  Working on a hydrostatic model of the photosphere of Arcturus by
treating atomic
opacity (continuous and line opacity) in non-LTE using large, model atoms, \citet{short:03} show
that overionization of Fe can cause important changes in the atmospheric structure.
Allowing for non-LTE,
the boundary temperature affecting the formation of water vapor was lowered, as inferred by \citet{ryde:02}.
This suggested that non-LTE effects in atoms may suffice to account for the
surprising presence of H$_2$O molecules in the photosphere of Arcturus \citep{ryde:02}.
Thus, relaxation of the assumptions behind classical model photospheres may
change the atmospheric structure. Even small alterations of
the heating or cooling terms in the energy equation (for example
due to dynamic processes or uncertainties and errors in the
calculations of radiative cooling) may lead to changes in the
temperature structure in the outer, tenuous regions of the
photosphere where the heat capacity per volume is low.

Finally, it should be noted that an elaborate model of the atmosphere of Betelgeuse also has to
accommodate an inhomogeneous chromosphere with hot and cool components \citep{tsuji_2000,harper:01}.
We do not attempt to achieve this level of sophistication in this paper.

Notwithstanding all these complications, we have based our nominal
stellar parameters for
modeling spectra of Betelgeuse on those of \cite{tsuji_2000}, namely
$T_\mathrm{eff}=3600$~K, $\log g= 0.0$, solar metallicity, and $M=15\,M_\odot$
(implying $R=646\,$R$_\odot$).
The temperature uncertainty from different investigations in the literature, with a span from 3140 to 3600 K or more should, however,
always be borne in mind.
Following \citet{tsuji_2000} the microturbulence is set to $\xi_\mathrm{micro}=4\,$\kms.
The metallicity is assumed to be approximately solar \citep{lambert:84,carr:00,lobel:00}, except for the C, N, and O abundances.
We choose these abundances as determined by \citet{lambert:84}, who based their analysis on molecular lines, an analysis which therefore
is sensitive to the assumed effective temperature. The C abundance does not change by large amounts with effective temperature, but
the derived abundances of oxygen and nitrogen do \citep[see Figure 6 in ][]{lambert:84}.
These authors determined the C abundance primarily from
vibration-rotation lines of CO. These are almost
independent of the oxygen abundance since all C is bound into CO throughout most of the atmosphere.
Thus, together with the C abundance from the CO lines, the OH lines can provide an estimate of the O abundance. Conversely,
the
strengths of the vibration-rotation lines of OH are sensitive to the O abundance but also to the C abundance, or
more correctly, to the (O-C) abundance. It should, however, be noted that
saturated lines are not as sensitive to the abundance as are weak lines.
For our nominal temperature of $T_\mathrm{eff}=3600$~K the values are
$A_\mathrm C = 8.29$\footnote{The abundance by number $N_\mathrm C$ is given through the following definition:
$A_\mathrm C\equiv \log N_\mathrm C - \log N_\mathrm H + 12$}, $A_\mathrm N = 8.37$, and $A_\mathrm O = 8.52$.
With these CNO abundances the (O-C) abundance is $1.6\times10^{-4}$,
which is close to the solar value of (O-C)$_\odot=2.0\times10^{-4}$.
\citet{lambert:84} used
a nominal effective temperature of $T_\mathrm{eff}=3800$ K
for which the derived abundances are $A_\mathrm C = 8.4$, $A_\mathrm N = 8.6$, and $A_\mathrm O = 8.8$
and therefore the (O-C) abundance is $3.8\times10^{-4}$, which is more than double (0.4 dex) the value
derived for 3600 K.

Using the model photosphere, we calculate synthetic spectra by
solving the radiative transfer in a spherical geometry. We
calculate the radiative transfer for points in the spectrum
separated by  $\Delta \tilde{\nu} \sim 1\,\mbox{km\,s$^{-1}$}$
(corresponding to a resolution of $\tilde{\nu}/\Delta \tilde{\nu}
\sim 300\,000$) even though the final resolution is lower.
Calculations in spherical symmetry will generally tend to decrease
the strengths of strong lines in the mid- and far-infrared.
In extreme cases even emission may appear.
The reason for this is
that an extended photosphere
will occupy a larger solid
angle at line wavelengths where the total opacity is large. This
will result in a larger flux at line centers and therefore weaker
lines than those resulting from calculations in plane-parallel
geometry. However, the extension of our
Betelgeuse models is small, but will
nevertheless affect the line depths of the water vapor and OH lines
which are studied here.\label{em}

To match the observed line widths, we introduce the customary artifice
of a macroturbulent broadening, with which we convolve our synthetic spectra.
This extra broadening also includes the instrumental profile and stellar rotation and does not
change the strengths of the lines. We find that we need a macroturbulence of
$v_\mathrm{macro}=12\,$\kms\ (FWHM).
We assume the turbulent velocities follow an isotropic
Gaussian distribution ($\varpropto\exp{-(v^2/(2\sigma^2)}$), which has a statistical standard
deviation of $\sigma = \mathrm{FWHM}/(2 \sqrt{2 \ln 2}) =
\mathrm{FWHM}/2.355 \approx 5\,$\kms.
Put another way the Gaussian convolution profile
($\varpropto\exp{-(v/v_\mathrm{D})^2}$) has a Doppler velocity (or most probable velocity, see \citet{gray:92})
of $v_\mathrm{D}=\mathrm{FWHM}/(2 \sqrt{\ln  2})=\mathrm{FWHM}/1.665\approx7\,$\kms. The latter value can, after correcting for the
instrumental profile and
stellar rotation, be compared to
the Gaussian dispersion found by \citet{gray:00}, namely $11\,$\kms.

In order to fit the modeled lines to the observed ones, a macroturbulent
broadening of approximately $12\,$\kms\ (FWHM) is required.  This
is significantly smaller than what is needed from fitting optical
and near-infrared lines \citep{gray:00,lobel:00},
but the same as used by \citet{jennings:86} and \citet{antares}
for their mid-infrared lines.
This difference might be expected since the mid-infrared lines are formed further out in
the photosphere due to the increase of opacity, and one might expect that the type of granulation
and mass-eruption motions would be smaller further out in the photosphere
(D. Gray, 2005, private communications).








In the discussion in Section \ref{disc} we will also include model MOLspheres in
addition to our `naked' stellar photospheres.
The isothermal, spherical MOLspheres are characterized by an inner and
an outer radius, a temperature, and a column density of water vapor.
In common with published MOLsphere models, the source function of
the MOLsphere is assumed to be constant and in LTE, i.e., given the Planck function for a
specified temperature, and the water-vapor density within the sphere is constant.

In our model the microturbulent velocity in the MOLsphere is specified, and
the synthetic spectra calculated
on the basis of
our model photospheres are used as an inner boundary condition. The impact parameters of the
rays are linearly distributed on the disk and between
the stellar radius and outer radius of the water shell.
The intensities thus calculated are
later integrated over the stellar disk and the rest of the sphere, resulting in a flux spectrum
of the MOLsphere with
our model photosphere as the underlying illumination source. The MOLsphere contains only water vapor
and we use the
line list of \citet{par}.

\section{THE OBSERVED SPECTRA}

Several rotational lines have been identified, and are marked in
the Figures, from the
ground and second vibrational states of the hydroxyl radical (OH) and
pure rotational-transitions of water vapor in the ground and
first excited vibrational states. In our synthetic spectra we also
see weak SiO and a few weak metal lines, but these are too weak to be
identified in the observed spectra.

\begin{figure}
\epsscale{1.00}
\plotone{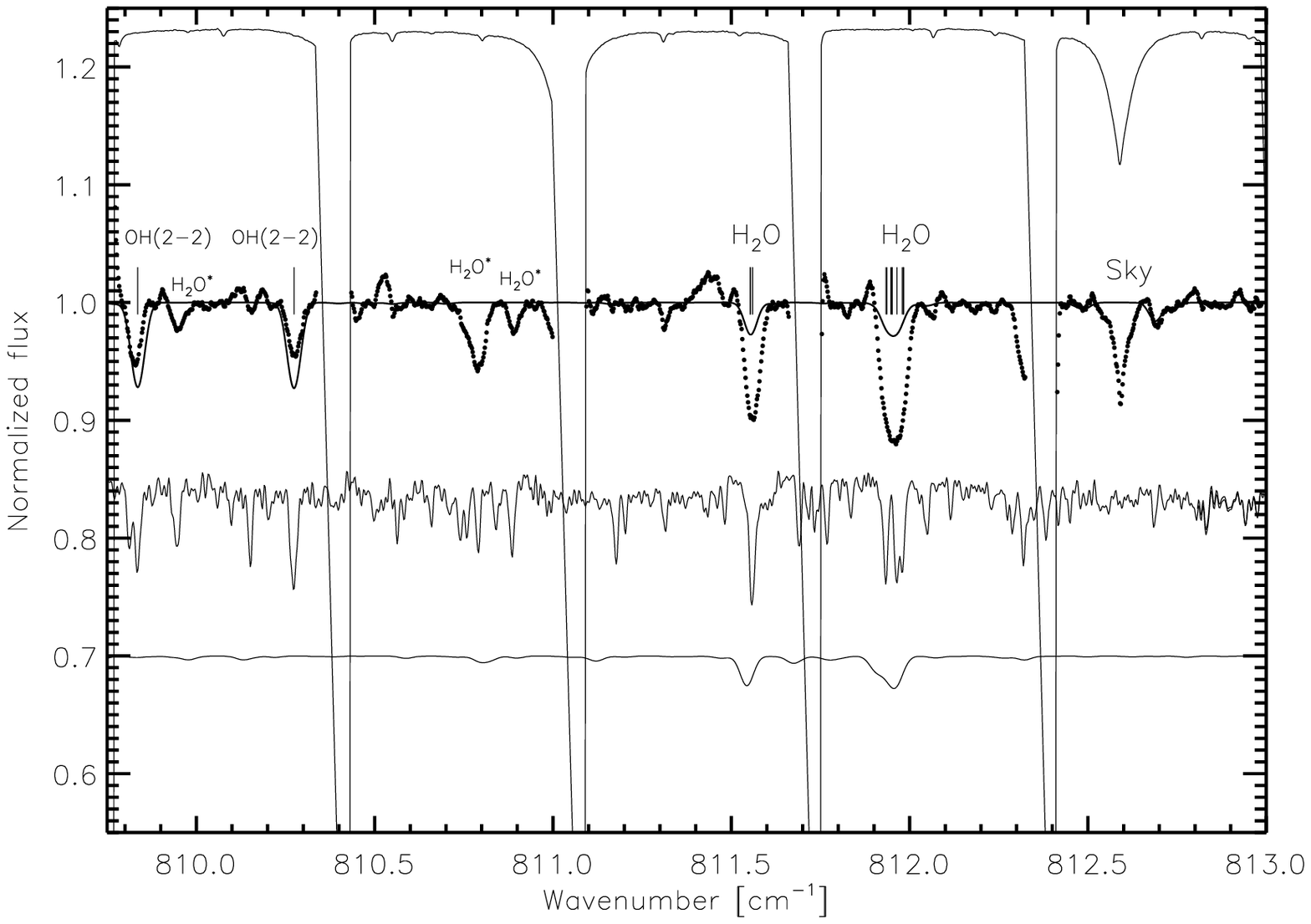}
\caption{High-resolution spectra of Betelgeuse in the $12\,\micron$ region, here $809.75-813.0\,$\invcm (dots), are
shown. The observations are corrected for the star's velocity at the time of observation. The wavelengths
are thus the laboratory wavelengths. Uppermost in the Figure, the sky spectrum is shown. The almost vertical
lines show the ranges of the spectral orders. In this figure five orders are present.
Overlaid on the observed Betelgeuse spectrum is a spectrum calculated on the basis of a classical, one-dimensional model
atmosphere ($\rm{T}_\mathrm{eff}=3600$ K)
with, most importantly, OH and water-vapor lines, but also weak metal lines (full line). Only a subset of water-vapor
lines in the wavelength region is
included. The water-vapor lines included are those with accurately measured wavelengths. A few other water-vapor lines are marked with an asterisk.
Shifted vertically downwards from the observed Betelgeuse spectra
is shown first the sunspot spectrum of this area. Next below is shown a pure water-vapor spectrum, based on the NASA-Ames
water-vapor line list for a Betelgeuse model.
Note that the NASA-Ames list includes a large number of lines but that the wavelengths are not
accurate enough for this resolution. Each OH and H$_2$O line included is indicated by a vertical
line, independent of their strengths.
 \label{obs_810_3600}}
\end{figure}

\begin{figure}
\epsscale{1.00}
\plotone{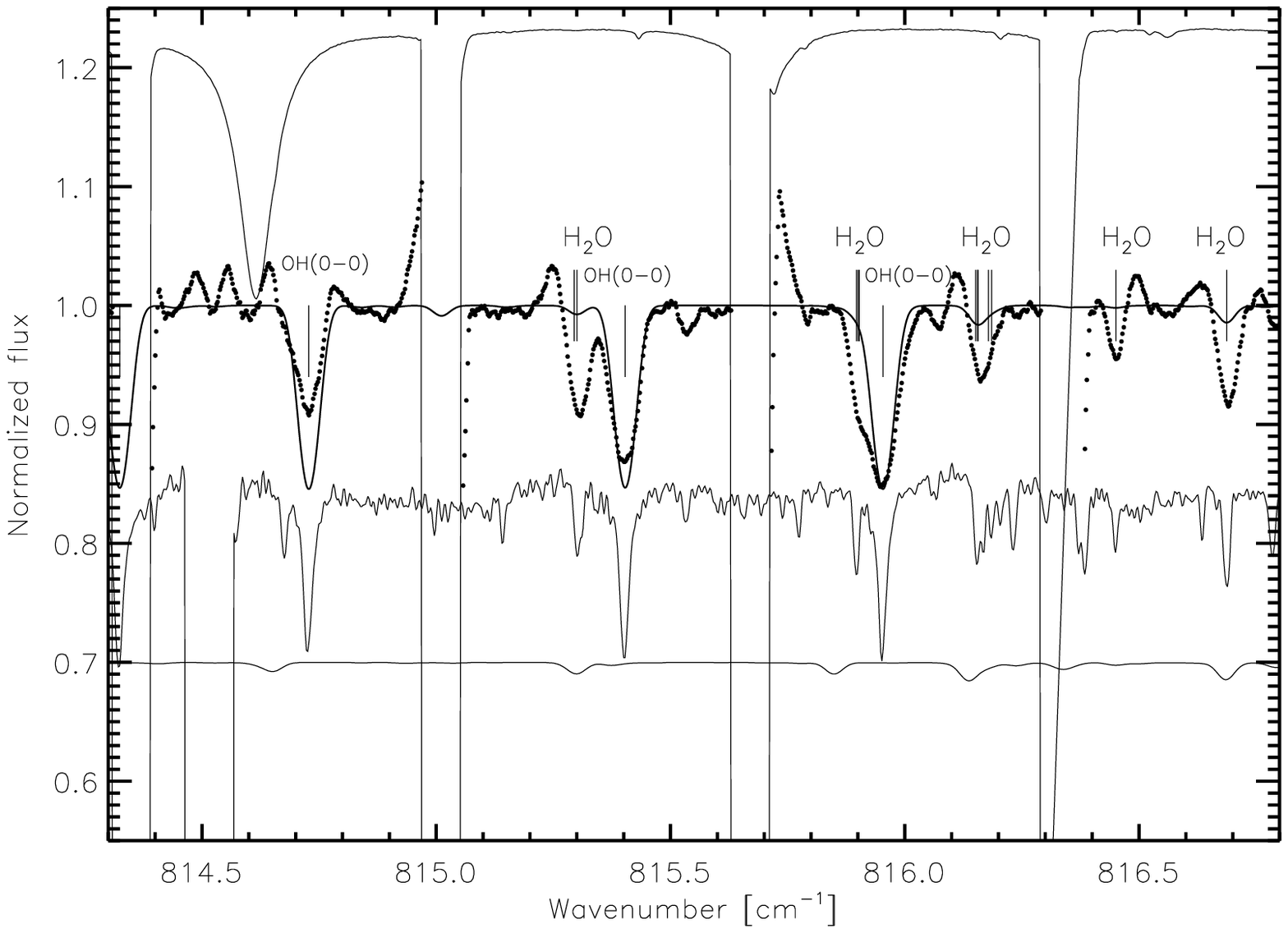}
\caption{For caption see Figure \ref{obs_810_3600}. The model spectrum shown is based on a model photosphere
of an effective temperature of $\rm{T}_\mathrm{eff}=3600$ K. Here the wavelength region $814.3-816.8\,$\invcm\ is shown.
 \label{obs_814_3600}}
\end{figure}

\begin{figure}
\epsscale{1.00}
\plotone{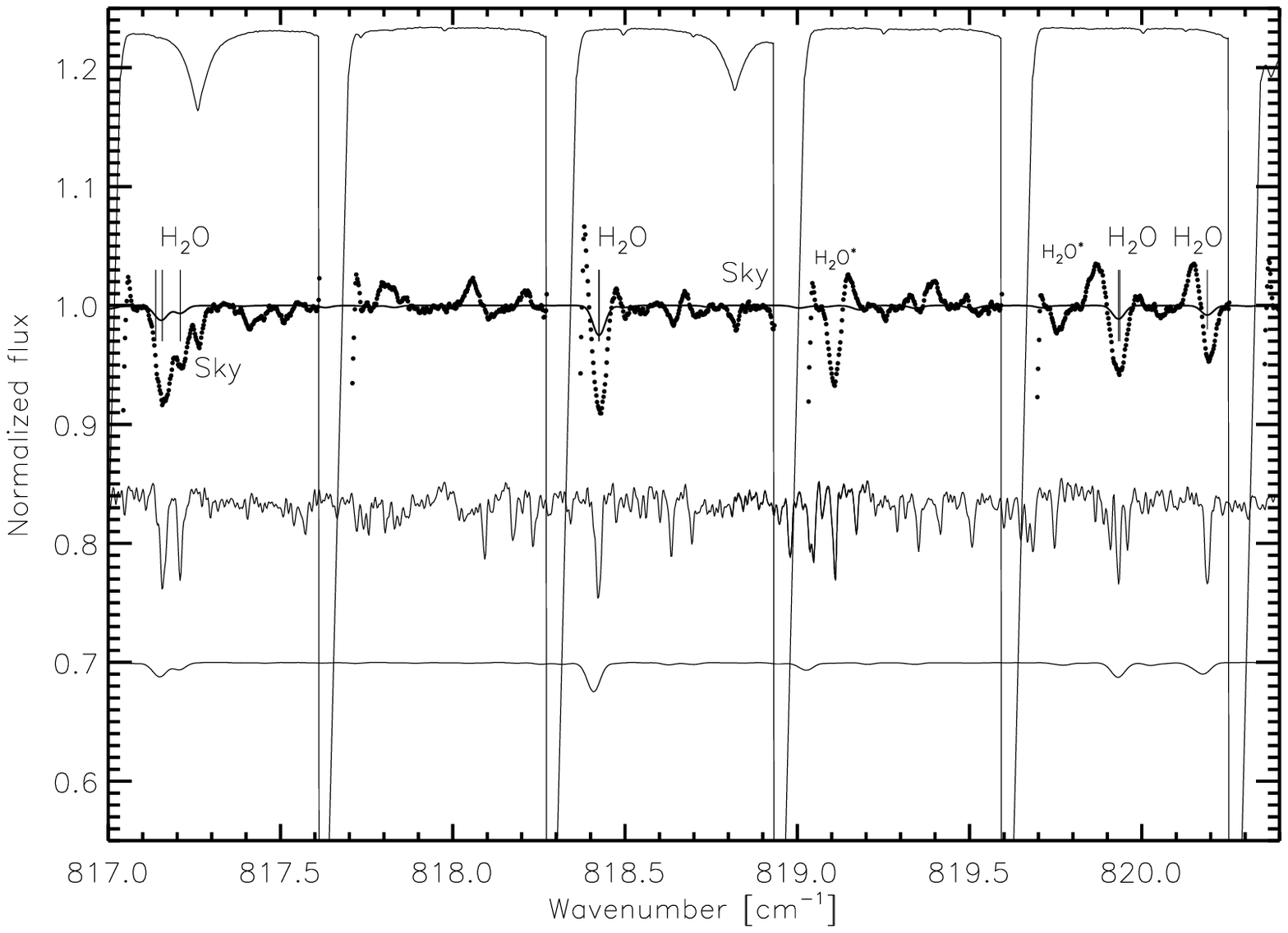}
\caption{For caption see Figure \ref{obs_810_3600}. The model spectrum shown is based on a model photosphere
of an effective temperature of $\rm{T}_\mathrm{eff}=3600$ K. Here the wavelength region $817.0-820.4\,$\invcm\ is shown.
 \label{obs_817_3600}}
\end{figure}

Overlaid on the observed spectra in Figures  \ref{obs_810_3600} - \ref{obs_817_3600} are calculated model spectra
based on a classical one-dimensional model atmosphere with an effective temperature of
$T_\mathrm{eff}=3600$ K. The line list used includes the metals and molecules that are relevant
in this wavelength region. The water-vapor lines cause some problems, however. The NASA-Ames'
water line list \citep{par} has a large number of lines and provides a complete water-vapor spectrum.
The wavelengths
of the list are, however, not accurate enough for high-resolution spectroscopy in the mid-infrared
as is discussed in \citet{ryde:02}. The uncertainty of the list is approximately
$0.05\,$\invcm\ or $20\,$\kms. Therefore, following
the procedure outlined in \citet{ryde:02}, only a subset of water-vapor
lines in the wavelength region is
included in the model spectra overlaid on the observations. The water-vapor
lines included are those with accurately measured laboratory wavelengths.  Each OH and H$_2$O line
included in the calculation is shown by a vertical
line, independent of their strengths.

The moderate mass loss of $\alpha$ Ori of approximately $4\times 10^{-6}\,\mathrm{M}_\odot$~yr$^{-1}$
\citep{glass} leads to an infrared excess, i.e. re-emitted thermal dust radiation.
We have estimated the contribution to the total recorded flux originating from this dust  as follows.
We estimate that our data have an effective
spatial resolution (a combination of seeing, diffraction, and guiding)
of slightly less than $2\arcsec$, which should be compared to our slit size of $2\arcsec.5\times 1\arcsec.5$. The specific intensity distribution
at $11\,$\micron\ was computed using the dust component model described in
Harper et al. (2001), which was based on fits to infrared interferometry data from Bester et al. (1996)\nocite{bester:96}
and \citet{sudol}. The specific intensity was convolved with the effective seeing,
and the resulting flux falling in the small TEXES slit calculated.

Following
\citet{ohnaka:04a}, our synthetic spectra are diluted by this extra,
optically-thin dust emission as follows:
\begin{equation}
F_\lambda^\mathrm{diluted}=(1-f_\mathrm{dust})\times F_\lambda+f_\mathrm{dust},
\end{equation}
\noindent where $F_\lambda^\mathrm{diluted}$ is the normalized modeled flux including the dust emission and $F_\lambda$
is the normalized synthetic spectrum. The attenuation of the photospheric light through the dust is
included in our definition of $f_\mathrm{dust}$.

We note, however, that assuming spatial resolutions of FWHM =1.\arcsec5, 1.\arcsec75, and 2.\arcsec0
for $11-12\,$\micron, we find that the dust dilution factor does not change by more than a few percent.
We arrive at a dust emission contribution to the total observed flux of
\begin{equation}
f^\mathrm{thin}_\mathrm{dust}=
(F_\mathrm{star+dust}- F_\mathrm{star})/(F_\mathrm{star+dust}) =
25\%
\end{equation}
\noindent for an optically-thin ($\tau \ll 1$), circumstellar dust envelope. The dust is indeed found
to be nearly optically thin in the mid-infrared, but there is, nevertheless, a small optical-depth effect
on the contribution of the dust radiation.
\citet{skinner} found a dust-shell optical depth at 9.7 \micron\ of $\tau=0.06$.
With the dust opacity used in our spectral modeling, we find that this value scales
to $\tau(12\,\micron)=0.04$.
Therefore, the estimated attenuation is $1-\exp(-0.04)=0.04$, a value we adopt.
Thus, if we take this attenuation of the photospheric light into account,
we arrive at a $f_\mathrm{dust} = 28\pm5\%$.
We note that \citet{schuller:04} found a smaller optical depth $\tau_{11\,\micron}=0.007$.
For their instrumental set-ups, \citet{antares} used an estimated
dust emission contribution of $f_\mathrm{dust}\sim 35\%$ at $12\,$\micron\ and \cite{ohnaka:04a}
used a value of $f_\mathrm{dust}=44\%$.


Under the observed Betelgeuse spectrum in Figures \ref{obs_810_3600} - \ref{obs_817_3600} the sunspot spectra
of this region are shown \citep{solspectrum}. With less macroturbulent broadening, these spectra clearly show
where water vapor and OH lines lie. \citet{wallace_science} reported on the
identification of numerous, resolved water-vapor lines in a
sunspot spectrum of the N-band ($760-1233\,\mbox{cm$^{-1}$}$).
\citet{poly_3} were later able to assign quantum numbers to these
lines, which is a very difficult task due to the great complexity
of the water-vapor spectrum.
The sunspots can be as cool as $3000\,\mbox{K}$ \citep{bernath_rev}
and Wallace et al. \nocite{wallace_science} derived an effective
temperature of the observed sunspot of approximately
$3300\,\mbox{K}$, which corresponds to a spectral class of early M \citep[see for example][]{allen}.
The figures also show a pure water-vapor spectrum, based on the complete NASA-Ames'
water line list \citep{par} for the  Betelgeuse model of $T_\mathrm{eff}=3600$ K.

A number of additional water-vapor lines seen in the
spectra of Betelgeuse and
the sunspot spectrum, but not calculated in the modeled spectra (which are overlaid on the observed spectra),
can clearly be identified. These water lines are marked above the observed spectra with an asterisk, i.e.
H$_2$O$^\star$, to indicate that they are not included in the modeled spectrum shown over
the observations. The wavelength shifts of these additional lines compared to the
corresponding ones in the Ames' list give an indication of this list's accuracy.

The upper-most spectra are sky transmission spectra. They show where strong telluric lines disturb the stellar spectra.
These lines are divided out but some residuals still remain. In some places, strong residuals are seen.
In each figure four or five spectral orders are shown, with gaps in the spectral
coverage due to incomplete
spectral coverage with TEXES beyond about 11 microns.

\section{ANALYSIS}

\subsection{WATER-VAPOR LINES}
\label{sect_h20}

In table \ref{H2O} our water-vapor line list is presented. Line positions with laboratory measurements from \citet{poly_1, poly_2}
are given with 8 digits (as is given in these references). The assignments are also from \citet{poly_1, poly_2}.
Following \citet{ryde:02}, other lines and blends of water-vapor lines not accounted for
in the laboratory measurements, are included from the \citet{par} line list.
These lines, given with only 6 digits, are included in our list with the (uncertain) wavelength
shifts found in the \citet{par} list. The $gf$-values, the
excitation potentials of the lower levels of the transitions, and the partition function are taken from \citet{par}.

In the table the measured equivalent widths and FWHM are also presented. Note, however, that many of the lines
are blended. Only five `truly single' lines are identified. The observed equivalent widths of the four spectral
features we have in
common with \citet{antares} all agree within $5-45\%$.
Furthermore, we note that when modeling our lines, we find that only the weak ones will have a FWHM given directly
from a convolution of the thermal broadening, microturbulence, macroturbulence,
and the instrumental profile, such as expected from lines in the weak-line approximation,
where the line optical depth is much smaller than the continuum optical depth, $\tau_l\ll\tau_c$
\citep[cf. for example ][]{gray:92}.
In our spectral synthesis, we have used a Gaussian microturbulent
velocity of $\xi_\mathrm{micro}=4\,$\kms\ (i.e., a Gaussian profile with a FWHM of  6.7 \kms)
which is constant throughout the atmosphere.
Furthermore, in the line-forming regions the temperatures are 2000-3000 K
which give thermal broadenings of $v_\mathrm{Doppler}=(2\mathrm{k} T/ m_\mathrm{water})^{1/2}=1.4-1.7\,$~\kms. This leads to a total
Gaussian velocity of 4.3 \kms\ or 7.1 \kms\ (FWHM).
With a macroturbulence of 12 \kms\ (FWHM) and a resolution of $R=70,000$ we arrive
at a width of 14.5 \kms\ as a modeled FWHM.

It is clear from Figures \ref{obs_810_3600} - \ref{obs_817_3600} that our nominal model atmosphere of
Betelgeuse does not represent a good fit to
the modeled water-vapor lines.
Given the concerns presented in Sect. \ref{models} of the adequacy of classical models, it is of interest
to investigate the effect of the temperature structure on the $12\,$\micron\ spectrum.

In Figures \ref{obs_810_3250} - \ref{obs_817_3250} the high-resolution spectra of Betelgeuse
are shown again, but now with a model comparison using a synthetic spectrum based
on a model atmosphere which is 350~K cooler, namely $T_\mathrm{mod}=3250$ K.
This temperature is actually close to the
temperature derived by \citet{bester:96} based on their size measurements at $11\,\mic$
and matching the total luminosity.
The C, N, and O abundances
have to be scaled to this new temperature, assuming that the IR lines used to determine these abundances also are formed in the
cooler environment, which might be questionable. We have scaled the derived abundances from \cite{lambert:84}
with -0.2, -0.3, and -0.4 dex for the C, N, and O abundances for the new temperature. This scaling was
guided by the discussion in \cite{lambert:84} and especially their Figure 6. Thus, the new C, N, and O abundances
are $A_\mathrm C = 8.08$, $A_\mathrm N = 8.05$, and $A_\mathrm O = 8.15$. These abundances
are highly uncertain, due to an uncertain extrapolation, and this fact should be borne in mind.

The synthetic spectra based on the cooler model show a much better fit to the observed water-vapor lines.
In Figures  \ref{obs_810_3250} - \ref{obs_817_3250}  the \emph{pure} water-vapor spectrum based on the NASA-Ames water-vapor
line list is also shown for a model photosphere of 3250 K. Recall that this list includes all important
water-vapor lines but does not provide sufficiently accurate wavelengths. However, the overall appearance of the water spectrum showing the influence of weak water lines and
the amount of continuum found is clearly demonstrated, and it is still in fairly good agreement with the
overall appearance of the observed spectra.



\clearpage
\begin{table*}
\footnotesize
\label{H2O}
  \caption{Observational data and the line list of the water-vapor lines.
}
  \label{water_12um}
  \begin{center}
  \begin{tabular}{lcccccc} \hline
  \noalign{\smallskip}
 \multicolumn{1}{c}{$\tilde{\nu}_\mathrm{lab}$}     &
   \multicolumn{1}{c}{$E''_\mathrm{exc}$} &
  \multicolumn{1}{c}{$\log gf$} &
  \multicolumn{1}{c}{$J'$$K'_a$$K'_c$$J''$$K''_a$$K''_c$}    &
  \multicolumn{1}{c}{$v_1v_2v_3$}&
  \multicolumn{1}{c}{Equivalent} &
  \multicolumn{1}{c}{FWHM}
  \\
   \multicolumn{1}{c}{ } &
  \multicolumn{1}{c}{} &
   \multicolumn{1}{c}{ } &
   \multicolumn{1}{c}{ } &
   \multicolumn{1}{c}{ } &
   \multicolumn{1}{c}{width } &
   \multicolumn{1}{c}{ }
    \\
   \multicolumn{1}{c}{[cm$^{-1}$]} &
   \multicolumn{1}{c}{[eV] } &
   \multicolumn{1}{c}{ } &
   \multicolumn{1}{c}{ } &
   \multicolumn{1}{c}{ } &
   \multicolumn{1}{c}{[$10^{-3}$\,cm$^{-1}$]} &
   \multicolumn{1}{c}{[km\,s$^{-1}$] }
  \\
 \noalign{\smallskip}
  \hline
 \noalign{\smallskip}
811.553\tablenotemark{b}    & 0.992 & -1.226  &  22  15  7    21  14    8     &  &  &    \\
811.55952\tablenotemark{b}  & 0.992 & -1.703  &  22  15  7    21  14    8 & (000) & 5.5\tablenotemark{a} &   17\tablenotemark{a}  \\
811.93309  & 1.166 & -1.229  &  21   15   6    20  14    7 & (010)  &  &   \\
811.936    & 0.983 & -1.785  &      & &    \\
811.948   & 1.166 & -1.702  &      &  &  \\
811.950  & 2.180 & -0.967  &   & &    \\
811.951   & 1.964 & -1.801  &     & &    \\
811.96458  & 0.983 & -1.308  & 23    13   10   22  12  11 & (000) &  10.0\tablenotemark{a}\tablenotetext{a}{This value is the total equivalent width for all lines included in the spectral feature identified as arising from water vapor.} &  26\tablenotemark{a} \\
811.97965  & 1.166 & -1.702  & 21    15   7   20  14    6 & (010) & & \\
811.982   & 2.180 & -0.967  &     &    \\
811.984    & 1.964 & -1.801  &     & &   \\
815.294    & 1.989 & -1.557  &     & &   \\
815.30059  & 0.498 & -2.51   & 18    7   12   17  4   13 & (000) & 5.7\tablenotemark{a} & 18\tablenotemark{a} \\
815.897\tablenotetext{b}{Uncertain assignment of the quantum numbers for the states of the transition.}\tablenotemark{b}
  & 1.396  & -1.00  & 21    21   0   20   20   1 & (010)&    \\
815.900\tablenotemark{b} & 1.396 & -1.48  & 21    21    0   20     20   1 & (010) & 1.1\tablenotemark{a} & 12\tablenotemark{a}  \\
815.903    & 3.237 & -0.744  &    &  &    \\
816.157    & 1.467 & -1.109  &     & &   \\
816.151    & 1.467 & -1.586  &       &  &   \\
816.15525  & 1.150 & -1.301  & 22    13 10   21  12   9  & (010)& 3.5\tablenotemark{a} & 18\tablenotemark{a}   \\
816.179    & 2.721 & -1.241  &       &&    \\
816.179    & 2.721 & -0.764  &      & &   \\
816.18551  & 1.150 & -1.778  & 22    13   9    21  12   10 & (010)&    \\
816.45026  & 0.398 & -3.21  & 17   5    13   16  2    14 & (000) & 2.3 & 13  \\
816.68703  & 1.014 & -1.35  & 24    12   13  23  11   12 & (000) & 4.9 & 18 \\
817.138    & 1.206 & -1.66  &     &  &   \\
817.15695  & 1.206 & -1.18  & 21    16   5    20  15   6  & (010) &  2.8\tablenotemark{a} &  15\tablenotemark{a}\\
817.20881  & 1.014 & -1.83  & 24    12 12   23  11   13 & (000) & 1.2 & 13  \\
817.209    & 1.460 & -1.55  &       &  \\
818.4238\tablenotemark{c}\tablenotetext{c}{Assignments from Tsuji (2000).}
           & 1.029 & -1.66  & 22    16  6    21  15    7 & (000)&    \\
818.4247\tablenotemark{c}    & 1.029  & -1.19  & 22    16   7    21  15    6 & (000) & 5.7\tablenotemark{d}\tablenotetext{d}{This value is uncertain since the line lies on an order edge.} & 16\tablenotemark{d} \\
819.93233  & 1.050 & -1.42   & 25    11   14   24 10  15 & (000)& 3.9\tablenotemark{a} & 18\tablenotemark{a} \\
819.937    & 1.918 & -1.455  &    &    &   \\
820.19016  & 1.360 & -1.039  & 21 20 1   20 19 2 & (010)  & 2.3 & 14 \\
\noalign{\smallskip}
  \hline
  \end{tabular}
  \end{center}
  \smallskip

\end{table*}
\clearpage
\subsection{OH LINES}

\begin{figure}
\epsscale{1.00}
\plotone{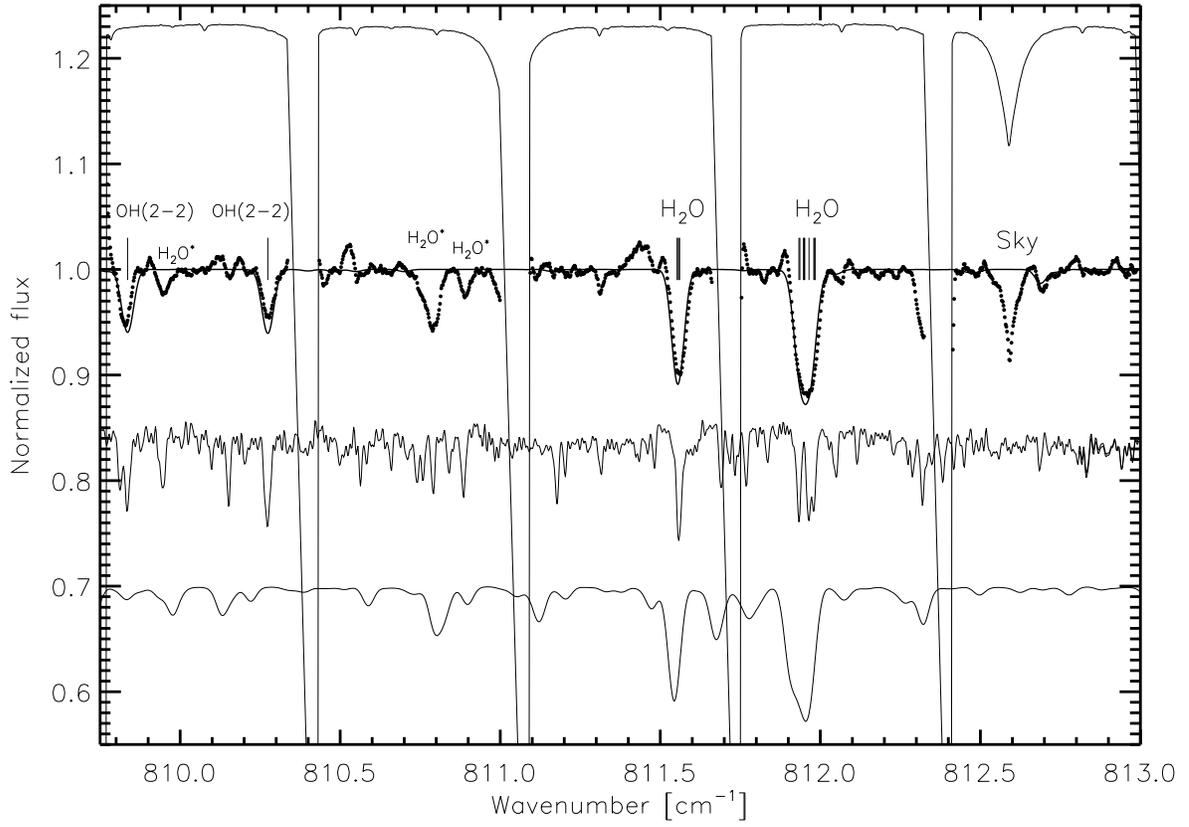}
\caption{These spectra show our $\alpha$ Ori observations in the same way as in Figure \ref{obs_810_3600}. However,
the model comparison is that based on a slightly cooler model photosphere ($T_\mathrm{mod} = 3250\,$~K).
 \label{obs_810_3250}}
\end{figure}

\begin{figure}
\epsscale{1.00}
\plotone{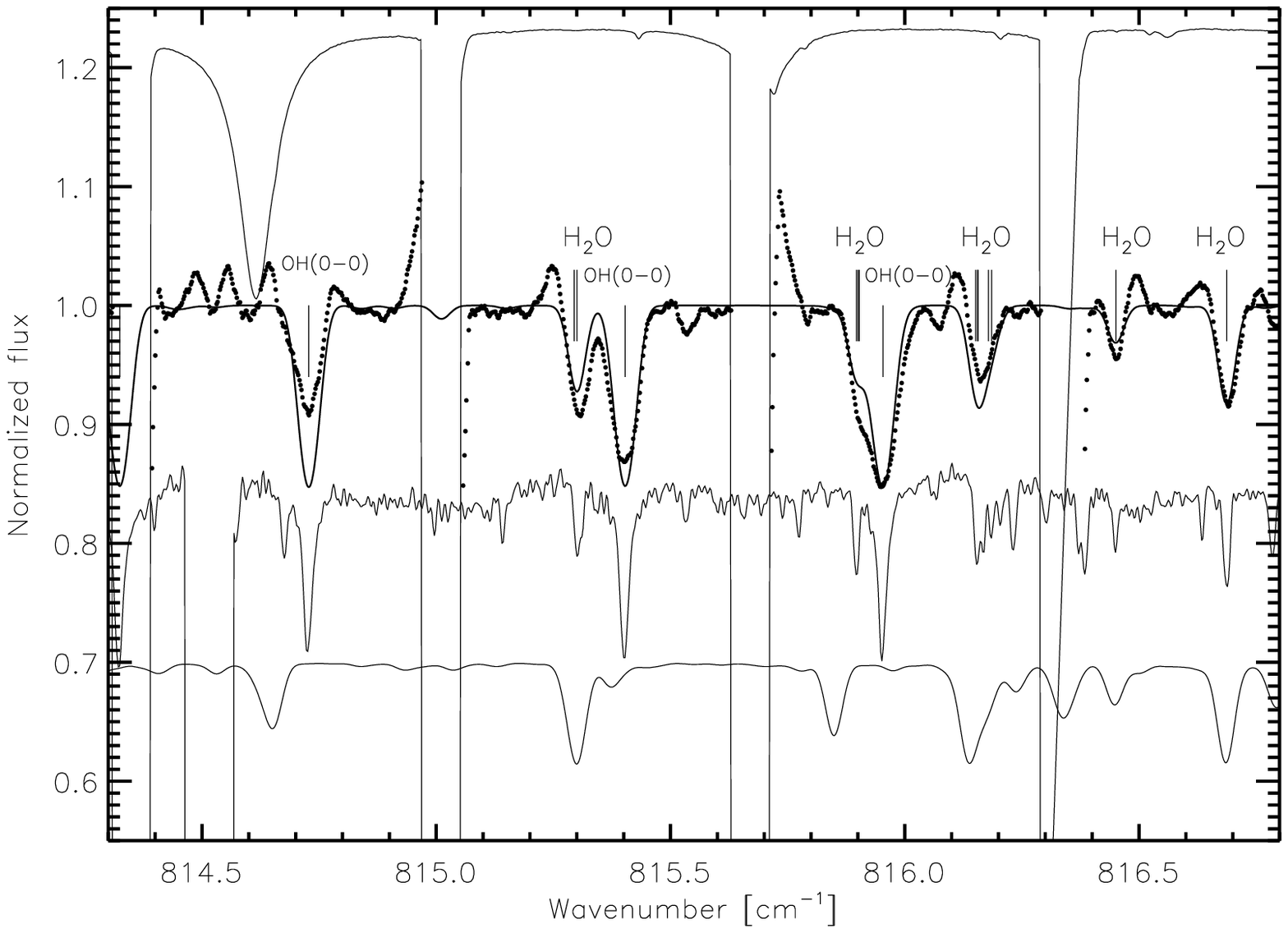}
\caption{For caption see Figure \ref{obs_810_3250}. The model spectrum shown is based on a model photosphere
of an effective temperature of $\rm{T}_\mathrm{mod}=3250$ K. Here the wavelength region $814.3-816.8\,$\invcm\ is shown.
 \label{obs_814_3250}}
\end{figure}

\begin{figure}
\epsscale{1.00}
\plotone{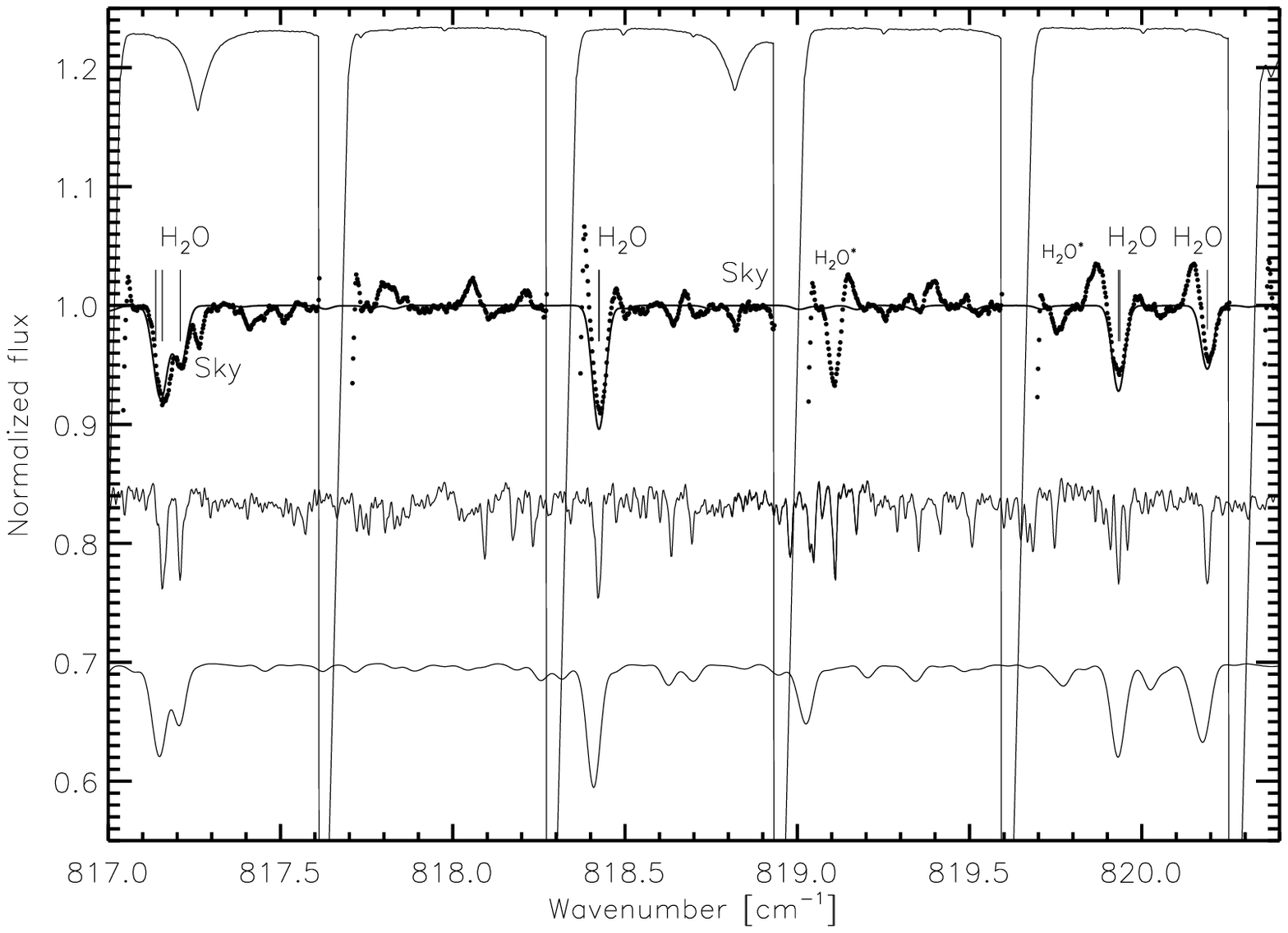}
\caption{For caption see Figure \ref{obs_810_3250}. The model spectrum shown is based on a model photosphere
of an effective temperature of $\rm{T}_\mathrm{mod}=3250$ K. Here the wavelength region $817.0-820.4\,$\invcm\ is shown.
 \label{obs_817_3250}}
\end{figure}

The OH lines come in quartets of intrinsically equal strengths. In figures \ref{obs_810_3600} and  \ref{obs_810_3250}
two OH($v=2-2$) lines of a quartet are shown. In figures \ref{obs_814_3600}  and \ref{obs_814_3250}
the four modeled lines in an OH($v=0-0$)-quartet are visible. Only three of the OH($v=0-0$) lines are, however,  observed, with the one at 814.7 \invcm\
appearing weaker as a result of the uncertain continuum fitting.  This
provides an estimate of these uncertainties.

In Table \ref{OH} the molecular data, measured and predicted
equivalent widths and FWHM are summarized. Laboratory line positions are
accurate to about $0.0004\,\mbox{cm$^{-1}$}$ or $0.15
\,\mbox{km\,s$^{-1}$}$ \citep{gold}. We use line data from the
\citet{gold} line list and the assignments in the Table are from
the mid-infrared Atlas of the sunspot spectrum
\citep{solspectrum}.
The predicted equivalent widths are calculated for our nominal
model with an effective temperature of $3600\,$K. The equivalent widths are measured, including the dust
contribution of $28\%$. Our predicted equivalent widths of the OH($v=0-0$) lines are within $5\%$ of
those calculated by \citet{tsuji_2000} (corrected for different amounts of dust contributions), who
used slightly
different CNO abundances and slightly larger oscillator strengths of the OH lines. These predicted widths are also in agreement with
the predictions for these lines made by \citet{jennings:86}.

We note that all three detected OH($v=0-0$) lines are, due to the intrinsically broad spectral lines in Betelgeuse, blended with water-vapor lines
to different degrees, which means that the measurements of their widths and equivalent widths
are subject to uncertainties.
The continuum level of the $R_\mathrm{1e}24.5$ line is not well defined, adding to the uncertainty of this line.
When determining the equivalent widths and FWHMs, we
have tried to disentangle the  blends affecting the  $R_\mathrm{2e}23.5$ and $R_\mathrm{1f}24.5$ lines by fitting two line profiles for the features,
thereby trying to avoid effects of obvious blends in
the measurement of the equivalent widths.

It is interesting
to note that the observed FWHMs of the OH($v=0-0$) lines are generally larger than
those of the water lines and
the OH($v=2-2$) lines, the latter being closer in value to the water-vapor lines. This
is true also for the modeled lines
and is a result of the saturation of the OH($v=0-0$) lines \citep[see also][]{jennings:86}. Since they are
saturated, they will also be sensitive to the microturbulence assumed.
Indeed, we find that the OH lines are quite sensitive to the microturbulence used in the calculation of the
synthetic spectra.
The water lines are not as sensitive.
When calculating the synthetic spectra with $\xi_\mathrm{micro}=3$, 4 (nominal), and $5\,$\kms, we find that
decreasing the microturbulence by $2\,$\kms\ (from $\xi_\mathrm{micro}=5\,$\kms\ to $3\,$\kms) decreases the equivalent
widths of the water lines by
$28\%$ and the OH($v=0-0$) lines by $45\%$.
The larger saturation of the OH lines increases this extra uncertainty of the microturbulence in the modeling
of the lines.

We find that the strengths of the modeled OH($v=0-0$) lines agree roughly with the observed strengths,
within the estimated uncertainties,
although the observed strengths
are larger and the uncertainties in the equivalent-width measurements tend to give too small values.
The rotational lines from the second vibrational excited state, OH($v=2-2$),
appear slightly too strong in the model.
The fits of the OH lines are, however, in general much better than those of the
water-vapor lines.
Synthetic spectra based on the model photosphere of a 350 K cooler effective temperature
show a better fit to the OH lines in general, see Figures \ref{obs_810_3250}-\ref{obs_817_3250}.

We note that the strengths of the observed OH lines are a factor of two to three
stronger than those presented in \citet{antares} and \citet{jennings:86}; the former find observed equivalent widths
of $W_{R_{2e}23.5}=3.4\pm1.5 \times 10^{-3}\,$\invcm\ and $W_{R_{1f}24.5}=3.4\pm1.0 \times 10^{-3}\,$\invcm.
This is in marked contrast to the relatively good agreement with the measured equivalent widths of the water-vapor lines in
Sect. \ref{sect_h20}. The \citet{antares} equivalent widths indicate that the OH lines are approximately half as
strong as the water-vapor lines at $811.6$ and $811.9$, and $815.3\,$\invcm, and equally as strong as the
water-vapor line at $818.4\,$\invcm, which is clearly not the case in our recorded data.
We do not know the reason for this discrepancy, but the higher-resolution of our data makes it much easier to
measure equivalent widths and to disentangle blends. Also, as noted earlier, the continuum-fitting in the infrared
is not a simple task in general and large uncertainties (individual in size for every line) may affect the lines.

Through their formation and dissociation in molecular equilibrium, the abundances of photospheric water vapor and OH depend
on the abundance of oxygen and on the temperature. In our 3600 K model, the abundance of OH is 100 times larger
than that of H$_2$O for the outer part of the photosphere ($\log\tau_\mathrm{Ross} < -1.5$) and with a rapidly
increasing ratio OH/H$_2$O further in. At the depths where the continuum is formed the ratio is approximately $10^3$.
In our 350 K cooler model (which has a lower oxygen abundance) the abundances of OH and H$_2$O increase, but the
water abundance increases by a factor of five more. From the continuum forming regions
and outwards for this model, the four most abundant molecules are, in decreasing order, H$_2$, CO, OH and H$_2$O.
To conclude, the water lines are much more sensitive to the
temperature than the OH lines, as can be seen in Figures \ref{obs_810_3600}-\ref{obs_817_3250}
where the two models shown differ by only 350 K.
The water-vapor lines are fitted much better after we lowered the
effective temperature by 350 K from our initial value of 3600 K
(and adjusting the C, N, and O abundances accordingly).

\clearpage
\begin{table*}
  \caption{Observational data of pure rotational, OH lines.
}
  \label{OH_2} \label{OH}
  \begin{center}
  \begin{tabular}{lccccccc} \hline
  \noalign{\smallskip}
 \multicolumn{1}{c}{$\tilde{\nu}_\mathrm{lab}$}     &
  \multicolumn{1}{c}{$E''_\mathrm{exc}$} &
  \multicolumn{1}{c}{$\log gf$} &
  \multicolumn{1}{c}{$v'-v''$}    &
  \multicolumn{1}{c}{Lower level} &
  \multicolumn{1}{c}{Obs. Equivalent} &
  \multicolumn{1}{c}{Pred. Equ.} &
  \multicolumn{1}{c}{FWHM\tablenotemark{a}}
  \\
   \multicolumn{1}{c}{ } &
   \multicolumn{1}{c}{ } &
   \multicolumn{1}{c}{ } &
   \multicolumn{1}{c}{ } &
   \multicolumn{1}{c}{ }&
   \multicolumn{1}{c}{Width\tablenotemark{a}\tablenotetext{a}{A colon (:) marks measured values with large
uncertainties.}} &
   \multicolumn{1}{c}{Width} &
   \multicolumn{1}{c}{ }
     \\
   \multicolumn{1}{c}{[cm$^{-1}$]} &
   \multicolumn{1}{c}{[eV] } &
   \multicolumn{1}{c}{ } &
   \multicolumn{1}{c}{ } &
   \multicolumn{1}{c}{ } &
   \multicolumn{1}{c}{[$10^{-3}$\,cm$^{-1}$]} &
   \multicolumn{1}{c}{[$10^{-3}$\,cm$^{-1}$]} &
   \multicolumn{1}{c}{[km\,s$^{-1}$] }

  \\
 \noalign{\smallskip}
  \hline
 \noalign{\smallskip}
809.8358 & 2.351 & -1.54 & 2-2 & R$_\mathrm{2e}26.5$ &  3:\tablenotetext{b}{This equivalent width is the measured value for our 3600 K model, including an extra dust emission of $28\%$.}  & 5.4\tablenotemark{b} & 15: \\ 
810.2738 & 2.349 & -1.52 & 2-2 & R$_\mathrm{1f}27.5$ &  2:  & 5.3\tablenotemark{b} & 14: \\ 
814.7280 & 1.297 & -1.57 & 0-0 & R$_\mathrm{1e}24.5$ &  7.4:  & 9.0\tablenotemark{b} & 23: \\ 
815.4032 & 1.302 & -1.58 & 0-0 & R$_\mathrm{2e}23.5$  &  10.4 & 8.9\tablenotemark{b} & 25\\ 
815.9535  & 1.300 & -1.57 & 0-0 & R$_\mathrm{1f}24.5$ &  12.9: & 9.1\tablenotemark{b} & 29:\\ 
\noalign{\smallskip}
  \hline
  \end{tabular}
  \end{center}
  \smallskip
\end{table*}
\clearpage

\section{DISCUSSION}
\label{disc}
We have presented clear water-vapor signatures in the M1-2 Ia-Iab supergiant star $\alpha$ Orionis.
At one time, water vapor was not expected
in the photospheres of giant
stars hotter than late M-type stars \citep[see, for example, the discussion in][]{tsuji_2001}.
However, based on \emph{ISO} data, \citet{tsuji_2001} presented evidence
of water-vapor signatures in Aldebaran ($\alpha$ Tau;
K5III) with an effective temperature of approximately 3900 K, and other cooler giants. These features were explained by introducing a MOLsphere.
Furthermore,
using high-resolution, mid-infrared spectra showing unexpectedly strong water-vapor lines,
\citet{ryde:02} presented evidence and argued for the presence of photospheric water
vapor in a star as early as Arcturus ($\alpha$ Bo\"{o}tis; K2IIIp) with an effective
temperature of approximately 4300 K. For Arcturus, there is no evidence of a MOLsphere \citep{tsuji:03}.
Instead, \citet{ryde:02} argued that the physical structure of the outer photosphere may be too crudely described:
A scenario in which the outer photospheric temperature
structure is not set purely by radiative-equilibrium or in which
non-LTE effects are important,
may be able to reproduce stronger water-vapor lines and to get the relative strengths
between near- and mid-infrared OH lines and mid-infrared H$_2$O lines correct.
Thus, it was possible to construct a semi-empirical, non-flux-conservative model photosphere that fitted
the OH and H$_2$O lines in the infrared wavelength region simultaneously. The reason this was
possible is the distinct response of the
water-vapor lines with respect to OH to the specific temperature structure of the outer photosphere and
the different height-of-formation of the lines in the photosphere. Note, however, that the suggested model for Arcturus presented in
\citet{ryde:02} would also produce strong TiO bands, which are not observed, if chemical equilibrium is assumed \citep[see the discussion in ][]{ryde:livermore}.

Our new, mid-infrared spectra of $\alpha$ Ori clearly reveal water vapor and the hydroxyl radical, OH.
The spectra are of such quality that close-lying lines can be resolved, which is significant in the analysis and
is an improvement compared to spectra of this wavelength region previously published.
The line strengths and widths can discriminate between possible models of the data.

We find that the strengths of the observed rotational lines of OH($v=0-0$), presented here, are in fair agreement with
synthetic spectra calculated on the basis of a standard,  one-dimensional, homogeneous, and hydrostatic model photosphere
of an effective temperature of 3600 K, an effective temperature often assumed for Betelgeuse.
However, we also show that the synthetic spectra fail to reproduce the observed rotational water-vapor lines,
in qualitative agreement with earlier studies:
the observed rotational lines of water vapor are too strong (see Figures \ref{obs_810_3600}-\ref{obs_817_3600}).

The interpretation of these lines is not simple. Our high-resolution spectra do, however,
give us some indications. The line widths are similar to photospheric lines,
which can be seen from Table \ref{H2O}, and later on from Figures \ref{obs_810_3250}-\ref{obs_817_3250}.
The photospheric velocity fields are quite large, much larger than observed in the inner S1 CO absorption
shell,
FWHM$\simeq 7\>{\rm km\>s}^{-1}$ \citep{bernat:79}, but also smaller than those observed in
optically-thin chromospheric UV emission features, FWHM$\sim 30\>{\rm km\>s}^{-1}$ \citep{carpenter}.
This suggests that the observed features are formed close to the star, perhaps in or near the photosphere,
but also we cannot rule out a close connection to the inhomogeneous chromosphere.

Another piece of information we can retrieve from our spectra is an upper limit of the excitation temperature. This will be useful later.
Assuming that the lines are not saturated and are formed in an isothermal thin layer in rotational equilibrium
close to the star, in parallel with the
interpretation of the $12\,\mic$ lines by \citet{antares}, the ratio of equivalent widths (W) of two lines (represented by the subscripts 1 and 2)
is given by
\begin{equation}
\frac{W_1}{W_2}= \frac{gf_1}{gf_2} \,e^{-(\chi_1-\chi_2)/kT},
\label{w}
\end{equation}
where $\chi$ is the excitation energy of the lower state of a transition and $T$ the temperature of the layer.
Thus, from the lines we have
detected we can calculate a mean temperature of such a layer, that is required from the observed lines strengths. If the stronger lines are saturated this
will be an upper limit of the temperature. We find
a temperature of $T=2600\pm200\footnote{standard deviation of the mean}$ K. This is in agreement with the upper limit of the temperature of
the assumed water-vapor layer, found by
\citet{antares}, namely $T< 2800$ K.
The temperature structure of our model photosphere of $T_\mathrm{eff}=3600$ K
reaches down to 2600 K first at $\log \tau_\mathrm{Ross}\sim -3.6$ which is far out.
The column density of water vapor through this model
is also quite low: $5\times10^{17}$\,cm$^{-2}$. This value should be compared with the estimate by \citet{antares}
of $(3\pm 2) \times 10^{18}$\,cm$^{-2}$. This explains the very weak water-vapor features in the synthetic spectrum based on our
model photosphere of 3600 K. As will be discussed later, we
find $N_\mathrm{col}=5\times10^{18}$\,cm$^{-2}$ for our model with 3250 K.

Thus, a synthetic spectrum based on a pure classical photosphere does not fit the observed spectra. What modifications could we make
to fit the lines? Here, we will start by discussing a few atmospheric models of Betelgeuse
which have recently been published.

\subsection{Can a MOLsphere explain our spectra?}

Molecular shells around giant and supergiant stars seem to be the emerging paradigm,
that has been put
forward in order to understand, first, the water-vapor signatures in the near-infrared
that cannot be explained by a classical model photosphere and, second,
the inconsistent interpretations of interferometric data. The question is whether such a layer could explain our new data of the
mid-infrared spectra of Betelgeuse.

In a homogeneous MOLsphere scenario (i.e., a layer surrounding the star with a constant density and temperature)
the intensity will be given, in LTE,  by $I_\nu\approx B_\nu(T_L)$ in an optically thick line and in the optically-thin case by
\begin{equation}
I_\nu\approx I_\nu(0)+[B_\nu(T_L)-I_\nu(0)] \,\,\tau_\nu^\mathrm{layer},
\end{equation}
where $I_\nu(0)$ is the specific intensity from the photosphere, $B_\nu(T_L)$ the Planck function of the MOLsphere layer
at temperature $T_L$, and $\tau_\nu^\mathrm{layer}$ is the optical depth through
a line-of-sight of the layer. Integrating over the disk of the star and the MOLsphere will give the flux.
Thus, the appearance of a
line, whether it is in absorption or in emission, will depend on the optical depth in the line
(which depends on the temperature and the column density of the layer)
and on the extension and temperature of the layer (through the Planck function).
The larger the layer, the more the emission, and
the temperature will also determine the relative strengths of the lines.

In the following models we have assumed a photospheric macroturbulent velocity
of $v_\mathrm{macro}=12\,$\kms\ (FWHM) as derived from observed spectra,
and a similar macroturbulent velocity for the MOLsphere. The assumption of the macroturbulent velocity in the
MOLsphere is based only on the need to fit
the line widths of the synthetic spectra to the observed line widths. Physically, it may be
more difficult to justify.
The total thermal and non-thermal broadening in the MOLsphere is assumed to be
$v_\mathrm{micro}=3\,$\kms\ (FWHM).

\begin{figure}
\epsscale{1.00}
\plotone{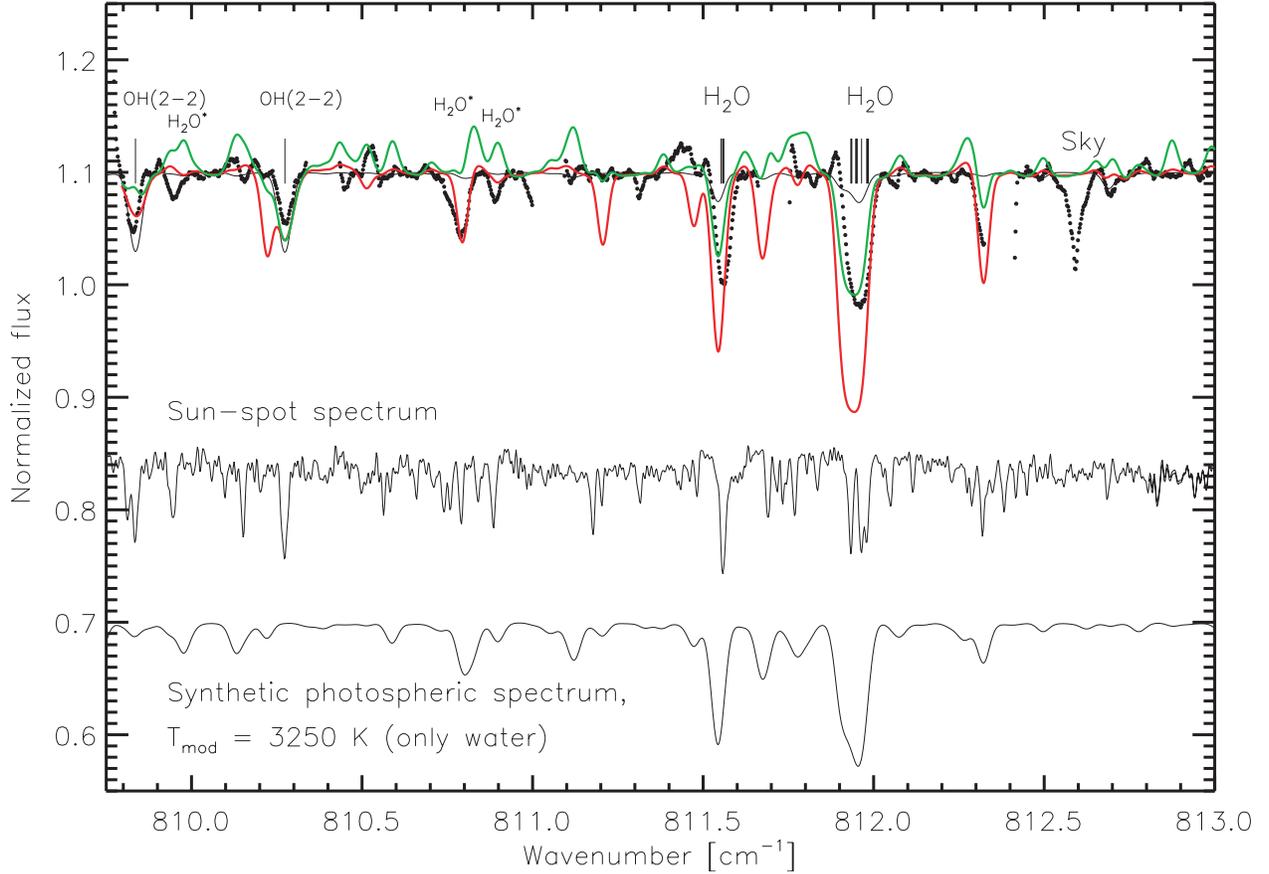}
\caption{
Two normalized spectra of the $809.8-813.0\,$\invcm\ region calculated for two MOLsphere realizations are shown by the green and red lines.
The MOLsphere is assumed to contain only water vapor at a temperature of $T=1500$ K and
has a column density of water vapor of $N_\mathrm{col}=10^{20}$\,cm$^{-2}$, as described in \citet{tsuji_2000}.
Absorption lines are formed but weakened by emission from an extended MOLsphere. We have modeled the layer extending from the
surface of the star to a radius of 1.45 (red) and 1.6 stellar radii (green). The more
extended the model, the more emission will `fill in' the absorption lines.
The underlying photospheric spectrum is shown by the full line normalized to one. Our observed TEXES spectrum is shown by dots.
For clarity we have shifted these normalized spectra up by 0.1. To be able
to make a comparison, we also show the sun-spot spectrum shifted
down to 0.85 and the normalized water-vapor spectrum of our cooler model photosphere shifted down to 0.7.
These are the same spectra also shown in Figure \ref{obs_810_3250}.
In all the synthesized spectra the water-vapor line list by \citet{par} is used and the dust emission contribution of 28\% is taken into account. All relevant lines (such as OH) are included in the
underlying photospheric spectrum.
It should be noted that the underlying photospheric spectrum is not identical to the one shown in Figure \ref{obs_810_3600},
the reason being the different line lists used. The water-vapor line list used here is more complete but
its wavelengths are not accurate enough for our high-resolution spectra. This explains that the lines do not match the observed lines perfectly.
The ordinate scale is the same as in Figures \ref{obs_810_3600}-\ref{obs_817_3250}.
\label{mol_t}
}

\end{figure}

\citet{tsuji_2000} interprets absorption features in the near-infrared as non-photospheric water vapor.
The MOLsphere needed to model these data
has a column density of water vapor of $N_\mathrm{col}=10^{20}$\,cm$^{-2}$
and a temperature of $T=1500\pm500$ K. The size of the MOLsphere is not constrained.
Due to the large column density of water-vapor and the low temperature needed to fit the near-infrared
bands of water vapor, deep absorption lines are
supposed to be formed but weakened by
emission from an extended MOLsphere \citep{tsuji_2000}.
We have modeled such a MOLsphere with our photospheric model as the underlying light source.
In Figure \ref{mol_t} we plot such a MOLsphere with two different extensions, namely a layer extending from the
surface of the star to a radius of 1.45 and one extending to 1.6 stellar radii. The more
extended the model, the more emission will `fill in' the absorption lines. The model extending out to
1.45 $R_\star$ overestimates the strong lines, but fits a few weak ones. As we extend the MOLsphere
out to 1.6 stellar radii, the two strongest water-vapor lines are well matched. However, a
majority of the lines are in emission, which is not seen in the observed spectra. The relative
strengths of the water lines are not reproduced. The line at approximately 810.8 \invcm\
is best fitted by the model extending out to 1.45 $R_\star$, but
we also note that a few lines, which are not seen in the observed spectra, appear in the
synthetic spectra based on the extension out to 1.45 $R_\star$. These are, for example,
lines lying at 810.2, 811.2, 814.55, and 819.45 \invcm. This indicates
that the temperature of the MOLsphere is not correct. The extension of the sphere
can be tuned to fit a few lines but not others.
Thus, this MOLsphere model, with $T=1500$~K, does not fit our spectra well.

\begin{figure}
\epsscale{1.00}
\plotone{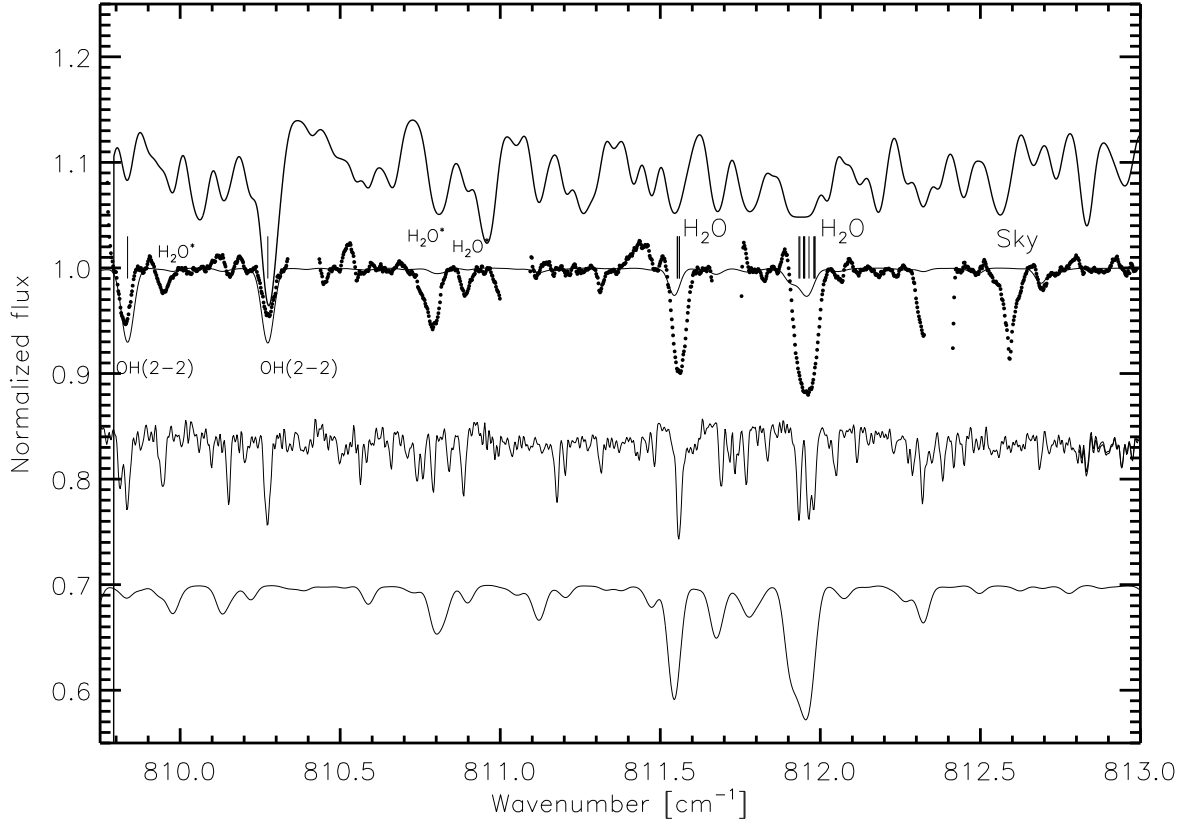}
\caption{The same figure as in Figure \ref{mol_t}, except that the realization of the MOLsphere is based on the one
suggested by \citet{ohnaka:04a}. The calculated MOLsphere spectrum with our underlying photospheric spectrum based on a
Betelgeuse model of 3600 K, is shown as the upper-most spectrum.
This spectrum is normalized to the photospheric spectrum, and is therefore mostly in emission,
due to the combination of its geometrical extension
the column density
of $N_\mathrm{col}=2\times 10^{20}$\,cm$^{-2}$
and temperature of of $T=2050$ K \citep{ohnaka:04a}. This MOLsphere extends
from the star out to a radius of 1.45 stellar radii.
Also in this figure, we show the sun-spot spectrum shifted down to 0.85 and
the normalized water-vapor spectrum of our cooler model photosphere
shifted down to 0.7.
 \label{mol_o}}
\end{figure}

A similar but warmer MOLsphere was modeled by \citet{ohnaka:04a}. This water-vapor layer surrounding
Betelgeuse was invoked in order to
explain partly mid-infrared, high-resolution spectra and partly the increasing diameter of Betelgeuse found by
interferometric measurements, from the K band to the $11\,\mic$ region. It was argued that the larger angular
size in the mid-infrared
originated from a MOLsphere containing water vapor.
The layer suggested by \citet{ohnaka:04a}, with water lines of optical thickness ranging from $\tau=0.1$ to $10$, also contributes
both absorption and emission,
due to its geometrical extension.
As was shown by \citet{ohnaka:04a}, his layer will not show strong water-vapor absorption
in the mid-infrared region, contrary to what we observe.
The column density
of this realization of a MOLsphere surrounding Betelgeuse is $N_\mathrm{col}=2\times 10^{20}$\,cm$^{-2}$
and the best fit to the data requires a temperature of $T=2050$ K \citep{ohnaka:04a}. It extends
from the star out to a radius of 1.45 stellar radii. In figure \ref{mol_o} the MOLsphere spectrum is shown together
with a part of our observations. The synthetic spectrum shows a large overall emission compared to the
photospheric continuum level. The synthetic spectrum consists mostly of a large number of emission lines that
blend into each
other. Therefore, setting the normalization level, for the comparison with the normalized observed spectrum, is difficult.
We have chosen to show the spectrum as calculated, normalized to the photospheric continuum.
Clearly the spectrum does not fit our high-resolution spectra, even if we allow for a different normalization.
Thus, based on our new data we can directly
rule out such a warm MOLsphere with a large column density surrounding Betelgeuse.
We note here that in order to explain their multi-wavelength, interferometric
observations of $\alpha$ Ori, \citet{perrin:04a} use an optically thick ($\tau_{11.15\mu\rm{m}}=2.3\pm0.2$)
molecular layer with a temperature of approximately $2050\,\rm{K}$ at $0.3\,\rm{R}_\star$ above the photosphere, similar to
the one argued for by \citet{ohnaka:04a}. We can, however, rule out such a scenario based on our high-resolution spectra.

\begin{figure}
\epsscale{1.00}
\plotone{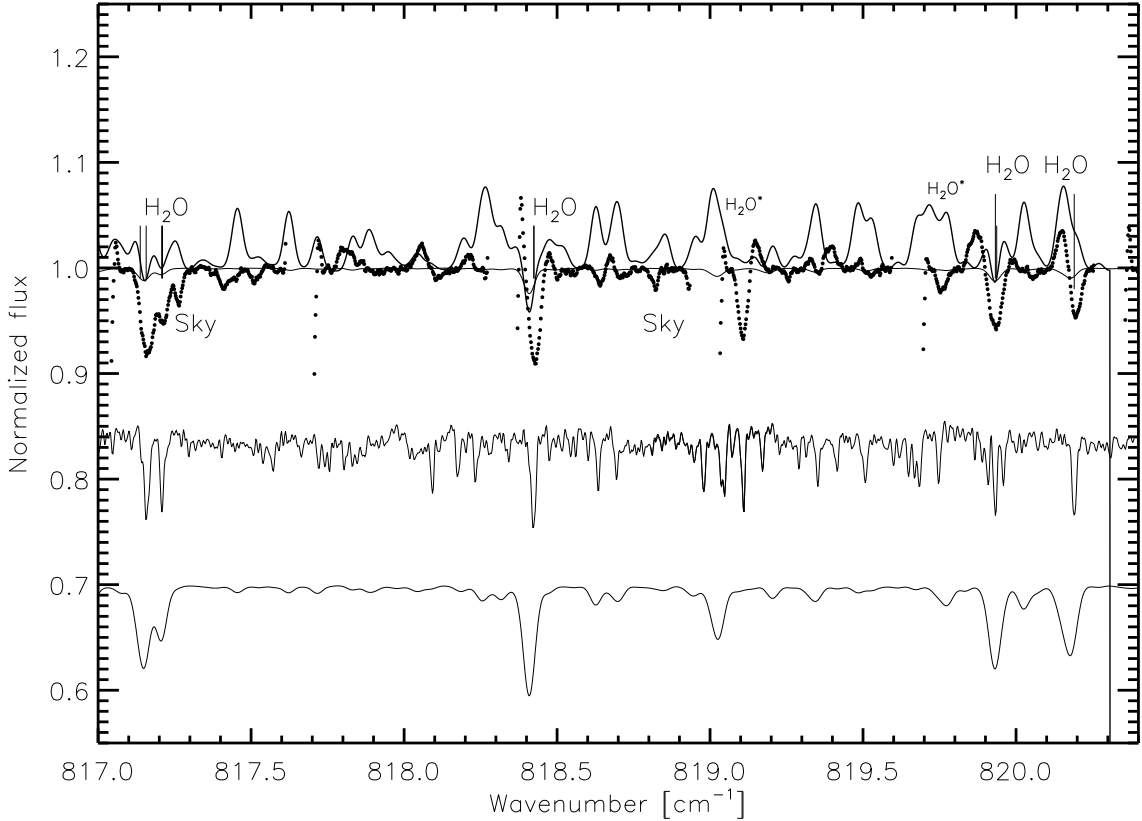}
\caption{In principle the same figure as in Figure \ref{mol_t}, except that the realization of the MOLsphere is that suggested by
Verhoelst et al. (accepted by A\&A). The wavelength region is different compared to Figure \ref{mol_t}. The emergent spectrum of the $817.0-820.4\,$\invcm\ region
is displayed by the thick line, mostly showing emission lines. The optically-thin layer of water vapor is
situated in a shell surrounding the star from approximately 1.3 to 1.45 stellar radii with a column density of
$N_\mathrm{col}=2\times 10^{19}$\,cm$^{-2}$ and a temperature of 1750 K. The dots shows our TEXES observations
and the spectrum shifted down to 0.7 is a normalized, pure water-vapor spectrum based on our cold Betelgeuse model of 3250 K and the
line list by \citet{par}.  The sun-spot spectrum shifted down to 0.85 is also shown. The underlying photospheric
spectrum is shown by the full line normalized to one.
\label{mol_v}}
\end{figure}

Recently, Verhoelst et al. (accepted by A\&A) modeled the extended atmospheric environment of Betelgeuse
in detail. To model both the near-infrared \emph{ISO} spectrum ($2.4-45\,\mic$) and infrared interferometric
information, they suggest the existence of an optically-thin layer of water vapor
situated in a shell surrounding the star from approximately 1.3 to 1.45 stellar radii with a column density of
$N_\mathrm{col}=2\times 10^{19}$\,cm$^{-2}$ and a temperature of 1750 K. This layer can explain
the overall appearance of the medium-resolution ($R\sim 1500$) \emph{ISO} spectrum. Furthermore, they
suggest that the size increase from
the K band to the mid-infrared can be explained by opacity from dust grains of alumina (Al$_2$O$_3$) instead of
the water-vapor layer suggested by \citet{ohnaka:04a}. Verhoelst et al. show that the column densities
suggested by \citet{ohnaka:04a} are too large to be compatible with the near-infrared data.
In Figure \ref{mol_v} we show the synthetic spectrum of the realization of the MOLsphere presented in Verhoelst et al. Due to
the extension of the MOLsphere and its temperature, the spectrum will, however, be dominated by emission lines.
Therefore, not even this MOLsphere realization fits our new data.

\begin{figure}
\epsscale{1.00}
\plotone{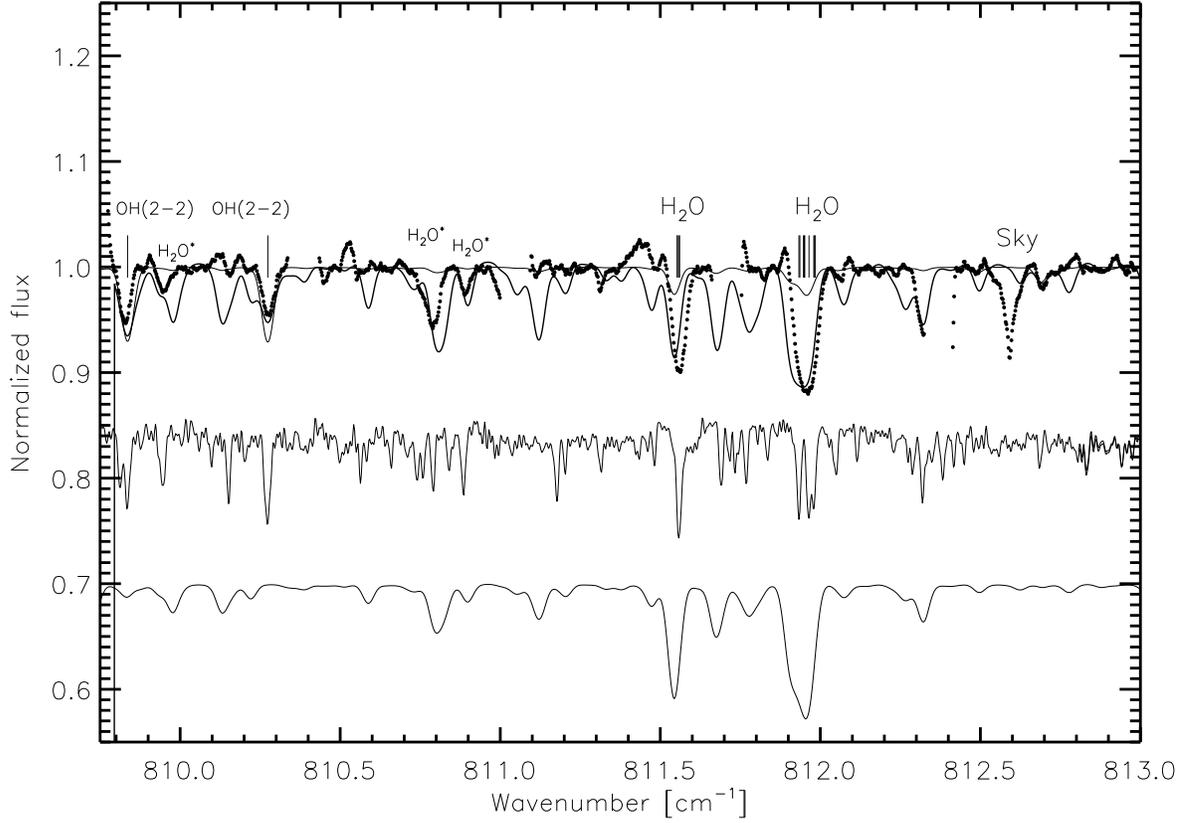}
\caption{
Similar figure as in Figures \ref{mol_t}-\ref{mol_v}. This MOLsphere realization simulates the case presented in \citet{antares}:
$N_\mathrm{col}=3\times 10^{18}$\,cm$^{-2}$ and a temperature of $T_L<2800$ K. The MOLsphere
extends from the stellar surface out to 1.01 stellar radii.
We find the best fit for a temperature of 2900 K. The MOLsphere spectrum is shown by a thick full line. The strong lines are
modeled well but several other lines are modeled too strong. The underlying photospheric spectrum is shown by the full line normalized to one.
The observations are shown by dots. The normalized spectrum shifted down to 0.7 is a pure water-vapor spectrum based on our cold Betelgeuse model of 3250 K and the
line list by \citet{par}. For comparison the sun spot spectrum is also shown.
 \label{mol_js}}
\end{figure}

\citet{antares} modeled their low-resolution, $12\,\mic$ spectra with a plane-parallel, isothermal
layer close
to the location of the onset of the chromospheric temperature
rise with $N_\mathrm{col}=3\times 10^{18}$\,cm$^{-2}$
and a temperature of $T_L<2800$ K. We have, therefore, simulated this scenario with a very thin MOLsphere on top of the photosphere. We have
assumed the MOLsphere to
extend from the stellar surface out to 1.01 stellar radii.
We find the best fit for a temperature 100 K warmer than the upper limit found by \citet{antares}, see Figure \ref{mol_js}.
We see that the strongest water lines are fitted well, but we also see that there might be several water-vapor lines
which are modeled too strong. We will discuss this model further in the next section.

As a summary, the realizations of a MOLsphere which have been presented in the literature in order to explain
signatures that can not be explained by classical model photospheres do not fit our high-resolution, mid-infrared spectra.
Further investigations of the $12\,\mic$ lines formed in a MOLsphere with an underlying photosphere
would be interesting.

\subsection{Can a cooler photosphere explain our spectra?}

Another way of looking at the water layer that \citet{antares} introduce in order to explain their spectra,
is that it represents a cooler outer photospheric structure where water is formed.
Thus, also inspired by the explanation of the unexpected water-vapor lines in the $12\,\mic$ spectrum of Arcturus
and the fact that the line widths and radial velocities of the water-vapor lines in $\alpha$ Ori
match that of the photosphere (similar to the case of Arcturus), we have investigated whether a cooling of the temperature structure
in the outer photosphere could even fit the observed $12\,\mic$ lines. The water-vapor
features, which are formed relatively far out in the photosphere, are very sensitive to the temperature where they are formed, much more so than the OH lines.
Given the complicating factors of the Betelgeuse photosphere compared to a one-dimensional, hydrostatic
model photosphere, accounted for in Section \ref{models}, it might not be surprising that a classical model
fails to reproduce spectra at all wavelengths.
Indeed, because OH and H$_2$O lines respond differently to temperature
variations,
a synthetic spectrum calculated from a photosphere of 3250 K (i.e. 350 K cooler),
resembles the observed spectrum much better (see Figures  \ref{obs_810_3250}-\ref{obs_817_3250}).
The fit of the water lines from the ground and first excited vibrational states is much better. The model
also predicts the OH lines better, especially OH($v=2-2$).
Note that this model is only used to simulate cool photospheric surface regions
(e.g. inhomogeneities) or a cooler
outer photospheric structure at the line-forming regions
of the observed lines.

It is interesting to note that the column density of water all the way through our cooler model photosphere
is $5 \times 10^{18}$\,cm$^{-2}$ which is close to what \citet{antares} found, namely
$(3\pm 2) \times 10^{18}$\,cm$^{-2}$. Also, our excitation temperature, which we estimated from Equation
\ref{w}, of $2600\pm200$ K is close to the temperature estimated by \citet{antares}. This temperature is
also reached deeper in our cooler model, at $\log\tau_\mathrm{Ross}\sim -2.2$. These facts
explain the good fit to our spectra.

Thus, it seems that our cool model, which can be seen as representing a colder temperature structure in the line-forming
regions, can explain the line widths (of the same order as the photospheric macroturbulence), relative line strengths,
and column density required. Our spectra are fitted well by our new model.

The effective temperature of our model photosphere that we need in order to model the spectra
is too low to be compatible with other spectroscopically (optically and near-infrared) determined effective temperatures of Betelgeuse.
The optical and near-infrared spectrum of Betelgeuse can  \emph{not} be modeled by a model photosphere of such a cold
temperature, see for example Verhoelst et al. (A\&A in press).
However, it is not unexpected that the outer photospheric structure could be colder, affecting only the mid-infrared region, in
analogy with the case of Arcturus \citep{ryde:02}, or
because the photosphere of Betelgeuse is inhomogeneous. Cool areas may dominate the spectral features in the mid-infrared
with the hot areas, in principle, only contributing extra continuous flux, depending on the temperatures and filling factors involved.
Measurements in the optical wavelength region are more biased towards hot spots or hotter areas, while cold areas
do not contribute much flux.
For example, a 3100\,K model has only $25\%$ lower flux in the mid-infrared, whereas in
the optical the flux drops by a factor of 30 as compared with a 3600\,K model.
Our exercise shows, that for the line-forming regions, a 3250 K model photosphere could be a better representation of
that part of the photosphere. The effect of this cooler outer structure on the near-infrared spectrum will have to be investigated.
The site of formation of the absorption lines at $12\,\mic$ could be a combination
 of both the photosphere and a MOLsphere.

Thus, it would be worthwhile to analyse the influence of
the physical structure of the outer photosphere on
the spectra, for example, by relaxing the assumption of LTE
in the calculation of the photospheric structure
\citep[see, for example, the case of Arcturus in][]{short:03}.
Furthermore, it would be worth investigating the effects on the lines introduced by photospheric inhomogeneities.
The observed water-vapor lines may also put constraints on 3-dimensional modeling of supergiants.
The challenge for any model including a more realistic
atmosphere including, a sophisticated photosphere and a MOLsphere, is that it should
not contradict any known, observed spectral features. We will investigate this possibility in a forthcoming paper.


\section{Conclusions}

We have observed high-resolution, mid-infrared spectra of Betelguese which we have demonstrated to be
difficult to explain in existing atmospheric scenarios. For example, a classical model photosphere is not able to explain our spectra, nor is
published MOLsphere models
which explain other observational discrepancies. We are, however, able to fit our observations with a synthetic
spectrum based on a cooler
(outer) photospheric temperature structure, as compared with a temperature structure calculated from a classical photospheric modelling.


Thus, the supposed non-classical and/or inhomogeneous outer
structures of the photospheres may
play a role and may have to be
taken into account as a complement to the emergent scenario of a MOLsphere, at least for supergiants.
It is clear that, when discussing the outer atmosphere, we require an understanding of the underlying photosphere
in its whole complexity, since it will be the underlying illumination
of any circumstellar matter. The classical modeling of the outer
photosphere of Betelgeuse is most probably also uncertain
given the neglect of the mechanical energy and momentum deposition that leads to
the chromospheric emission and mass loss.

Further investigations have to be
made to reconcile the different emerging pictures of the puzzling atmosphere
of Betelgeuse with the aim to be able to explain its entire spectrum in one united scenario.
Clearly, it is essential that all aspects and complexities of the photosphere must be investigated, including inhomogeneities,
non-standard photospheric structures, etc.

One of the characteristics of the putative MOLspheres is the
presence of envelopes which either have large density scale-heights, or discrete shells, overlying
the photosphere, which in turn has relatively small density scale-heights.
Interferometry with the Atacama Large Millimeter Array (ALMA) may be able
to test for the presence of the large density scale-heights or narrow shells associated with
these MOLspheres. Centimeter radio continuum interferometry with the Vary Large Array
has already been used to map the density scale-heights in the
inner wind and chromosphere \citep{harper:01},
and at 7~mm (43~GHz) the angular diameter of Betelgeuse is 90~mas \citep{lim:98}.
This is larger than the size of the MOLspheres, however, and similar measurements
with ALMA at higher frequencies, i.e., 100-700~GHz, will resolve the angular
diameter of the atmosphere and probe deeper in towards the photosphere.
Measurements of the angular diameter as a function of
frequency will provide a direct test of the density structure and of the existence
of MOLspheres.

\begin{acknowledgements}
NR would like to thank B. Plez and E. Josselin for fruitful discussions and their
hospitality during his stay at GRAAL in Montpellier. We gratefully acknowledge the suggestions made
by Bengt Gustafsson and Kjell Eriksson for improvements to the paper.
We are grateful for the help
of the TEXES team as well as the IRTF staff.
The use of TEXES is supported by
the National Science Foundation (NSF) grant AST-0205518
and by the Texas Advanced Research Program.
NR received some financial support from the Swedish Research Council and
GMH's contribution was supported by the NASA ADP Grant No. NNG04GD33G issued
through the Office
of Space Science and the NSF US-Sweden Cooperative Research Program grant INT-0318835 to the
University of Colorado. MJR received some financial support from the NSF (grant AST-0307497) and NASA (grant
NNG04GG92G).  TG received support from the Lunar and Planetary Institute, which
is operated by the Universities Space Research Association under NASA
CAN-NCC5-679. JL was supported by NSF grant AST-0205518.
\end{acknowledgements}

\end{document}